%Paper: hep-th/9509040
%From: SAIDI EL HASSANE <saidi@ictp.trieste.it>
%Date: Fri, 8 Sep 1995 12:02:51 +0200 (MET DST)

\hfuzz=15pt
\def\bbbz{\kern 1pt\lower 1pt\hbox{${\rm z}\!\!\!{\rm z}$}}
\ICIR{181}
\vskip-0.7truecm
\TITLE{On the$ N=2 U(1)$ supercovariant Lax formalism and W\hat
A(n-1\vert n-1)^{(1)}$symmetries.}
\vskip0.8truecm
\AUTHOR{E.H. Saidi}
\centerline{International Centre for Theoretical Physics, Trieste, Italy}
\centerline{and}
\centerline{Section de Physique des Hautes Energies Facult\'e des
Sciences, LMPHE,}
\centerline{Av. Ibn Batota, B.P. 1014, Rabat, Morocco.}
\vskip0.8truecm
\ABSTRACT

We introduce the concept of conformal spin gradation of the untwisted
affine Lie superalgebra $\hat A(n-1\vert n-1)^{(1)}$ to study the
$W\hat A (n-1\vert n-1)^{(1)}$ Miura transformation. We show that
the essential of $\hat A(n-1\vert n-1)^{(1)}$ may be read from the
conformal spin gradation of the canonical vector basis of the $SL(n\vert
n)$ vector representation space $V_{2n}$ and a spectral parameter
$\mu$. We give the generic formula of their conformal spin weights.
Then, we set up the fundamentals of a manifestly $N=2\ U(1)$ Lax formalism
leading to a manifestly $N=2\ W\hat A(n-1\vert n-1)^{(1)}$ Miura
transformation. Its explicit form is obtained and is shown to have a
similar structure as in the $N=0$ case. Both $N=0\ W\hat A(n-1)^{(1)}$
and $N=2\ W\hat A(n-1\vert n-1)^{(1)}$ Miura transformations
involve $(n-1)\ N=0$ and $N=2$ conserved currents with integer
conformal spins. The leading cases are discussed. Using the $U(1)$ charge
of the $N=2$ algebra, we develop also a new method of constructing $N=2$
superfield realizations of the $N=2$ higher spin supercurrents. Among other
results, we find that in general there are three series of $(n-1)$ higher
conformal spin $N=2$ supercurrents. The usual $N=2$ super $W$ currents
are the only hermitian ones. At the $n=3$ level, we find a new Feigin Fuchs
type extension of the conformal spin one $N=2$ supercurrent. Such a
feature, which has no analogue at the $n=2$ level, is also present
for $n>3$. Finally, we give the $N=2$ superfield formulation of the
$N=2$ Boussinesq equation and its generalization involving complex
$N=2$ supercurrents.

\MIRAMARE
\vfill\eject

\SECTION{INTRODUCTION}

The Virasoro algebra and its supersymmetric extensions together with
the Kac--Moody symmetries have played a central role in the study of
string dynamics and two dimensional conformal field theories (CFT).
These are symmetries generated by conformal spin $s$ currents, with
$s\leq 2$.

Some years ago, Zamolodchikov opened a new issue of possible extensions
of the Virasoro algebra to higher conformal spins. Among his results,
the discovery of a new extension, $W_3$, involving, besides the usual
spin two conformal current, a conformal spin three current [1]. This algebraic
structure is, however, a nonlinear Lie algebra.

Recently there has been some interest in studying these higher
conformal spin extensions of the Virasoro algebra by using different
methods: geometrical [2], Lie algebraic [3, 4, 5, 6], or also field theoretical
techniques [7, 8]. Progress in gauging these symmetries was made [9].

$W$ algebras and their supersymmetric extensions are known to arise
from the second Hamiltonian structure of the generalized KdV hierarchy
[10, 3, 4, 5]. They have then the property of relating integrable systems
and CFT models [11, 8]. Super $W$ algebras are also intimately
related to the affine Lie superalgebras. In the $\hat A_{n-1}^{(1)}$ case
for example, there are $(n-1)$ conserved bosonic current $\omega_s$
of conformal spin $s$ with $2\leq s\leq n$. The free field realizations of
these currents are given by the $W\hat A_{n-1}^{(1)}$ Miura
transformation [3, 9]. The conformal spin one current $\omega_1$ is
forbidden by the traceless condition of the $n\times n\ \ \widehat{SL}
(n)^{(1)}$ matrix representation. In the case of the untwisted affine Lie
superalgebra $\hat A(n-1\vert n-1)^{(1)}$ there are $(n-1)\ N=2$
supermultiplets $J_k$, with $k$ integer belonging to the set
$\{ 1,\dots n-1\}$, of conserved currents, see Eqs.(1.3) and (4.23--25).
Note that contrary to the bosonic case, it is the conformal spin $n$
supercurrent $J_n$ which vanishes. The spin one $N=2$ supermultiplet
is non zero. It is just the usual $N=2\ U(1)$ super Virasoro current.
This feature is not a coincidence and should have an explanation and
especially if one knows that $N=0$ and $N=2$ CFT's share many basic
properties [12, 13]. The understanding of this behaviour will give much
insight on $N=2\ U(1)$ integrable theories and probably $N=4\ SU(2)$
ones [14]. It allows, for instance, the derivation of the manifestly $N=2$
supersymmetric $W\hat A(n-1\vert n-1)^{(1)}$ Miura transformation
leading in turn to the manifestly $N=2$ superfield realizations of the
$N=2$ $W$ supercurrents. To our knowledge such a typical formula
has not been written down in the literature. All known $W\hat A(n-1\vert
n-1)^{(1)}$ Miura transformations are based upon breaking $N=2$
supersymmetry down to $N=1$ [15, 16].

In this paper we develop the manifestly covariant $N=2 U(1)$
supersymmetric Lax formalism and study the $W\hat A(n-1\vert n-1)^{(1)}$
symmetries and their implications in affine integrable $N=2 U(1)$
supersymmetric theories. Our main results are summarized as follows:

 First, we show that the essential of the $2n\times 2n\ \hat A(n-1\vert
 n-1)^{(1)}$ matrix representation can be read from the conformal spin
 gradation of the $2n$ canonical vector basis $\vert a>, a=1,\dots 2n$, of
 the $SL(n\vert n)$ vector representation space $V_{2n}$. The generic
 formula of these conformal spin weights (CSW) reads as:
 $$\vbox{\def\tablerule{\noalign{\hrule}}
\offinterlineskip\baselineskip=16pt\halign{\strut#&
\vrule#&
\hfil#\hfil&\vrule#&\hfil#\hfil&\vrule#&\hfil#\hfil
&\vrule#&\hfil#\hfil&\vrule#\cr

&\omit&&&$\vert k>$&&\quad $\vert k+n>$\quad&\omit\cr
\tablerule
&\omit&\quad CSW\quad&&\quad $n+{1\over 2}-k$\quad
&&$n-k$&\omit\cr}} \ ,
 \eqno(1.1)$$
 where $1\leq k\leq n$. The CSW of the dual vectors have the opposite
 values. We show also that $\Lambda_+$, the sum of conformal spin

 ${1\over 2}\ \hat A(n-1\vert n-1)^{(1)}$ generators satisfy a set of
 remarkable properties, namely
 $$\eqalign{
   \Lambda_+ &=\Lambda^+_++\Lambda^-_+\cr
   (\Lambda^+_+)^2 &=(\Lambda^-_+)^2=0\cr
   \{\Lambda^+_+,\Lambda^-_+\} &=\lambda_{++}\oplus
   \lambda_{++}\cr
   [\Lambda^+_+,\Lambda^-_+] &=\lambda_{++}\ominus\lambda_{++}\ ,\cr}

                   \eqno(1.2)$$
 where $\lambda_{++}$ is nothing but the sum of conformal spin one
 $\widehat{SL}(n)^{(1)}$ generators. These equations play a crucial role
 in the setting of the $N=2\ U(1)$ Lax formalism.

 Second, we give a detailed analysis of the $W\hat A(n-1\vert n-1)$ Miura
 transformation leading to $N=2$ superfield realizations of the $N=2$
 super $W$ currents. We review the $\hat A(n-1\vert n-1)$ Lax
 formalism using $N=1$ superfields and show that the $2(n-1)\ N=1$
 real supercurrent $U_\ell$, $2\leq \ell\leq 2n-1$, can be cast into $(n-1)$
 real $N=2$ supercurrents $J_k$ as:
 $$\eqalign{
   J_k &= U_{2k}\oplus U_{2k+1},\quad 1\leq k\leq n-1\cr
   &=\left(\omega_k,\omega^\pm_{k+{1\over 2}},\omega_{k+1}
   \right)\ .\cr}

                \eqno(1.3)$$
 This structure shows that the $N=2$ superfield analysis of
 $\hat A(n-1\vert n-1)^{(1)}$ may be compared with the $N=0$ bosonic
 treatment of $\hat A(n-1)^{(1)}$. Using this feature and the properties
 of $\Lambda_+$ Eqs.(1.2), we develop a manifestly
 $N=2\ U(1)$ Lax
 formalism. The $N=2\ U(1)$ Lax superoperators are
 $${\cal D}^\pm_+ =D^\pm_++Q^\pm_++G^\pm_+\ ,                \eqno(1.4)$$
 where $D^\pm_+$ are the usual $N=2$ covariant derivatives,
 $Q^\pm_+$ are $N=2$ complex conjugate spinor superfields and $G^+_++
 G^-_+=\Lambda_+$. The main result here is the derivation of a
 manifestly
 $N=2\ W\hat A(n-1\vert n-1)^{(1)}$ Miura transformation
 namely:
 $$\left(\nabla^n_{++}+\sum^{n-1}_1\ T_k\ \nabla^{n-k}_{++}\right) =
   \left(\mathop{\Pi}\limits^{n-1}_{j=1}\ \left(\nabla_{++}-J_{++}
   \right)_j\right)\ \nabla_{++}\ ,
\eqno(1.5)$$
 where
 $$\eqalign{
   \nabla_{++} &= [D^+_+,D^-_+]\cr
   <j\vert J_{++}\vert j> &= D^+_+Q^-_{+j}-D^-_+Q^+_{+j}+
   2Q^+_{+j}Q^-_{+j}\ .\cr}
        \eqno(1.6)$$
 This formula has been obtained by imposing the following constraints
 on the Lax operators Eqs.(1.4):
 $$\eqalign{
   \{{\cal D}^+_+,{\cal D}^+_+\} &= \{{\cal D}^-_+,{\cal D}^-_+\} =0\cr
   [{\cal D}^+_+,{\cal D}^-_+] &= {\cal Z}^{[+-]}_{++}\ ,\cr}
\eqno(1.7)$$
 where $Z^{[+-]}_{++}$ is a conformal spin one object valued in the
 $\left[\widehat{gl}(n)^{(1)}/gl(1)\right]\oplus\left[
 \widehat{gl}(n)^{(1)}/gl(1)\right]$ affine Lie algebra. Whereas the
 first constraints seem natural, the second one exhibits the similarity with
 the $N=0$ bosonic $\widehat{SL}(n)^{(1)}$ case. Note that no analogue
 object exist in the $N=1$ case but could exist for higher $N$'s.

 Moreover, we develop a new method of constructing $N=2$ superfield
 realization of the $N=2$ $W$ supercurrents from the $N=1$ superfield
 $W\hat A(n-1\vert n-1)^{(1)}$ Miura transformation by using a
 technique based on the $U(1)$ charge of the $N=2$ superalgebra.
 We show that all the $U'_k$ supercurrent split into five blocks as
 shown below
 $$\eqalign{
   U_{2k} &=(\kappa )^k\ J^0_k+J^{++}_k+J^{--}_k
   \quad\  1\leq k\leq n-1\cr
   U_{2k+1} &= (\kappa )^k\left(\Gamma^+_{k+1}+
   \Gamma^-_{k+{1\over 2}}\right),
   \quad 1\leq k\leq n-1\cr}
\eqno(1.8)$$
 where $\kappa$ is a real constant which is determined by the normalization
 condition of $J^0_1$. Higher $U(1)$ charge representations are
 forbidden by $N=2$ chirality and the Grassmann structure of fermions.
 $J^0_k$ are shown to describe the usual $N=2$ hermitian $W$
 supercurrents. The $n=2$ and $n=3$ examples are studied with details.
 Among other results we find a new Feigin Fuchs type extension of the
 conformal spin one $N=2$ supercurrent, namely
 $$J^{[+-]}_{++}=J_{++}(\psi^\pm_+)+J_{++}(\highchi^\pm_+)+
   \highgamma_0\highchi^-_+\psi^+_--\bar{\highgamma}_0\highchi^+_-
   \psi^+_+\ ,
          \eqno(1.9)$$
 where $\psi^+_+$ and $\highchi^+_+$ are fermionic $N=2$ chiral
 superfields and where $J^{[+-]}_{+-}(\phi^\pm_+)$ is the usual $N=2$
 superfield current, see Eqs.(1.6), (6.17) and (7.25). We obtain also the
 $N=2$ superfield realizations of the $\hat A(1\vert 1)^{(1)}$ and
 $\hat A(2\vert 2)^{(1)}\ N=2$ supercurrents.

 Finally, we study the $N=2$ supersymmetric extension of the
 Boussinesq equation and derive a generalization involving $N=2$
 complex supercurrents.

 The presentation is as follows.

 In Sections 2 and 3, we study the untwisted affine $\hat A(n-1\vert
 n-1)^{(1)}$ Lie superalgebra by using the concept of conformal spin
 gradation. Section 2 deals with the $n=2$ and $n=3$ cases, a matter of
 illustrating the idea and Section 3 describes the generic $n$ case. The
 essential of the $\hat A(n-1\vert n-1)^{(1)}$ may be read from Eq.(1.1)
 and the properties of $\Lambda^\pm_+$. In Section 4, we give the full
 analysis of the $W\hat A(n-1\vert n-1)^{(1)}$ Miura transformation.
 First we review the $N=0$ and 1 Lax formalisms. Then we set up the
 fundamentals of the manifestly $N=2\ U(1)$ Lax method and derive
 the manifestly $N=2\ W\hat A(n-1\vert n-1)$ Miura transformation.
 Two illustrating examples are given. In Section 5, we describe the
 $U(1)$ charge decomposition method of the $U_{2k}$ and $U_{2k+1}$
 supercurrents. In Sections 6 and 7, we give the complete study of the
 $U(1)$ charge representations for the $\hat A(1\vert 1)^{(1)}$ and\break
 $\hat A(2\vert 2)^{(1)}$ Lie superalgebra, respectively. In Section 8,
 we discuss the $N=2$ superfield formulation of the Boussinesq
 equation and give a generalization involving $N=2$ complex
 supercurrents. Our conclusion is given in Section 9. In Appendices A
 and B, we collect some useful formulas.

 \SECTION{AFFINE LIE SUPERALGEBRA $\hat A(m\vert m)^{(1)}$}

 In this section, we review briefly the basic properties of the untwisted
 affine Lie superalgebra $\hat g[m]=\hat A(m\vert m)^{(1)},  m\geq 1$
 which will be needed in this study. Also, we introduce the concept of
 conformal spin gradation of the purely fermionic simple root system of
 $\hat g$. This concept turns out to play an important role in the
 construction of the $W_s$ supercurrents and also in the study of
 integrable generalized KdV hierarchies. Moreover, as $\hat g[m]=g[m]
 \otimes\bbc [z,z^{-1}]$ where $g[m]=A(m\vert m)=SL(m+1\vert m+1)/<c>$,
 $c$ is the one--dimensional centre of $SL(m+1\vert m+1)$, we shall
 use the $2n\times 2n$ supermatrix representation of $SL(m+1\vert m+1)$
 to study this affine superalgebra. The centre $c$ is then given by the
 supertraceless $2n\times 2n$ unit matrix. Furthermore, in order to
 illustrate the idea of conformal spin gradation of $\hat g[m]$ and its
 implications on simple examples, we shall start by describing the main
 lines of the standard approach of $\hat g(1)$. Then we study the
 $\hat g(2)$ case by using the conformal spin gradation technique.
 We examine also the crucial role played by the hermitian form of the
 $SL(3\vert 3)$ vector representation space $V_6$. The generalization of
 these results for any $m>0$ is considered in the next section.

 \SUBSECTION{The $\hat A(1\vert 1)^{(1)}$ affine Lie superalgebra}

 First of all recall that the Lie superalgebra $SL(2\vert 2)$ has rank 3
 and can be realized in terms of the $4\times 4$ square matrices having
 a vanishing supertrace. Its dimension $d$ is equal to 7+8, i.e. generated
 7 bosonic generators and 8 fermionic ones. Contrary to the standard Lie
 algebras, $SL(2\vert 2)$ has three inequivalent simple root systems and
 thus three Dynkin diagrams namely: [17, 18]

$$\eqalign{
  \bigcirc\hbox to 30pt{\hrulefill}\bigcirc\hbox to 30pt{\hrulefill}
  \bigcirc\qquad\qquad {\rm (a)}\cr
  \bigcirc\hbox to 30pt{\hrulefill}\bigcirc\hbox to 30pt{\hrulefill}
  \bigcirc\qquad\qquad {\rm (b)}\cr
  \bigcirc\hbox to 30pt{\hrulefill}\bigcirc\hbox to 30pt{\hrulefill}
  \bigcirc\qquad\qquad {\rm (c)}\cr}
\eqno(2.1)$$
 where the white dot $\bigcirc$ and the black one $\bigcirc$ represent
 a bosonic and a fermionic simple root respectively. As required by
 supersymmetry [19, 8], we take the purely fermionic root system
 $\{\alpha_i\ \  i=1,2,3\}$ whose Cartan matrix reads as:
 $$K_{ij}=\left(\matrix{ 0&1&0\cr -1&0&1\cr 0&-1&0\cr}\right)\ .

              \eqno(2.2)$$
 Let $\{ E_{\pm\alpha_i}, H_i\}$ be the generators in the chevalley basis,
 associated with the fermionic simple root system $\{\alpha_i\}$,
 obeying
 $$\{ E_{+\alpha_i},E_{-\alpha_j}\}=\delta_{ij}\ H_i\ .
\eqno(2.3)$$
 Then one can express the remaining 4 even and 2 odd generators
 corresponding to the non simple roots as:
 $$\eqalign{
   (E_{++})_{12} &=\{ E_{+1},E_{+2}\}\cr
   (E_{++})_{23} &=\{ E_{+2},E_{+3}\}\cr
   (E_{+++})_{123} &=[E_{+1},(E_{++})_{23}]=[(E_{++})_{12},
   E_{+3}]\ ,\cr}
          \eqno(2.4)$$
 together with
 $$\eqalign{
   (E_{--})_{12} &= (E_{++})^+_{12}\cr
   (E_{--})_{23} &= (E_{++})^+_{23}\cr
   (E_{---})_{123} &=(E_{+++})^+_{123}\ ,\cr}
\eqno(2.5)$$
 where we have used the convention notation $E_{\pm\alpha_i}=E_{\pm i}$.
 All the other supercommutation relations are easily deduced
 with the help of
 Eqs.(2.3). Note that even generators carry an even number of charges
 whereas fermionic ones carry an odd number. Note also that the Cartan
 matrix Eq.(2.2) is degenerate because $SL(2\vert 2)$ has a non trivial
 centre $c=H_1+H_3=I_{4\times 4}$. Factorizing out this centre, one gets the
 simple Lie superalgebra $A(1\vert 1)=SL(2\vert 2)/<c>$. Its rank and
 its dimension are respectively 2 and $6+8$.

 The untwisted affine extension of $A(1\vert 1)$ denoted hereafter by
 $\hat g[1]=\hat A(1\vert 1)^{(1)}$ is defined in the usual way as
 $$\hat g[1]=g[1]\otimes\bbc [z;z^{-1}]\oplus\bbc\ d\ ,         \eqno(2.6)$$
 where $d$ is the derivation of $\hat g$ which induces the following
 natural $\bbz$--gradation of $\hat g$ [20].
 $$\hat g=\mathop{\oplus}\limits_{k\in\bbbz}\ \hat g_k         \eqno(2.7)$$
 with
 $$[d,\hat g_k]=k\ \hat g_k\ .
  \eqno(2.8)$$
 This particular $\bbz$--gradation is known as the homogeneous
 gradation of $\hat g$. We shall introduce another $\bbz$--gradation of
 $\hat g$ later on. Before that, recall that $\widehat{SL}(2\vert 2)^{(1)}$
 has two inequivalent simple root systems encoded in the following
 Dynkin diagrams
 $$\matrix{
   &&\bigcirc&&\cr
   &&\cr
   \bigcirc&&\bigcirc&&\bigcirc\cr}
\eqno(2.9)$$
 \bigskip
 $$\matrix{
   &&\bigcirc&&\cr
   &&\cr
   \bigcirc&&\bigcirc&&\bigcirc\ .\cr}
\eqno(2.10)$$
 We take the purely fermionic simple root system $\{\alpha_a,\
 a=(i,4)\}$ whose Cartan matrix reads as [14].
 $$K_{ab}=\left(\matrix{
   0&1&0&-1\cr
   -1&0&1&0\cr
   0&-1&0&1\cr
   1&0&-1&0\cr}\right)\ .
\eqno(2.11)$$
 The generators of this simple root system in the chevalley basis are
 given by $\{ (E_{\pm i},H_i), (E_{\pm 4}, H_4)\}$. The $(E_{\pm i}, H_i)$'s
 are the same as in Eqs.(2.3)--(2.5) and the generators $E_{\pm 4}$
 associated with the highest root (HR) $\alpha_4$ read as:
 $$\eqalign{
   (E_+)_4 &=\mu^2_{++}(E_{---})_{123}\cr
   (E_-)_4 &=\mu^{-2}_{++}(E_{+++})_{123}\cr
   H_4 &=\{E_{+4},E_{-4}\}\ ,\cr}
\eqno(2.12)$$
 where $\mu_{++}$ is a spectral parameter behaving as the inverse of
 the two--dimensional space--time coordinate $x_{--}=z$. In type I
 hierarchy [20], one requires:
 $$0=\partial_{++}\ \mu_{++}\equiv\partial_z\mu_z\equiv\partial\mu\ .

               \eqno(2.13)$$
 Recall that a similar constraint appears in the Lax formulation of
 bosonic integrable theories [3]. Note also that the homogeneous degrees
 $h$ of $(E_{\pm a},H_a)$ read as:
 $$\eqalign{
   h(E_{\pm i}) &=h(H_a)=0\cr
   h(E_{\pm 4}) &=\pm\ 2\ . \cr}                        \eqno(2.14)$$
 Moreover, in order to put on the same level the fermionic structure of
 the generators of $\hat A(1\vert 1)^{(1)}$ with the two--dimensional
 space--time spinors, we introduce the concept of conformal spin gradation
 of $\hat g$ [1] by asserting to its basic generators $E_{\pm i}$ and to
 the spectral parameter $\mu_{++}$ the following conformal spin weights
 (CSW):
 $$\eqalign{
   {\rm CSW} (E_{\pm i}) &=\pm\ {1\over 2}\cr
   {\rm CSW} (\mu_{++}) &= {\rm CSW}(\partial_z)=1\ ,\cr}   \eqno(2.15)$$
 so that the conformal weights of $E_{\pm 4}$ read as:
 $$\eqalign{
   {\rm CSW} (E_{+4}) &=2-{3\over 2}={1\over 2}\cr
   {\rm CSW} (E_{-4}) &=-2+{3\over 2}=-{1\over 2}\ .\cr}     \eqno(2.16)$$
 Thus we have the new $\bbz$--gradation of $\hat g$
 $$\hat g=\mathop{\oplus}\limits_{k\in\bbbz}\
   \hat g^{k/2}
     \eqno(2.17)$$
 with ${\rm CSW}\left(\hat g^{k/2}\right)={k\over 2}$. This is the
 conformal spin gradation of $\hat g$. In the next subsection and
 Section 3, we shall give more details on this conformal gradation and its
 implications in $N=2$ conformal and affine Toda theories. For the time
 being let us end this paragraph by recalling that the Cartan $h$,
nilpotent $\eta$ and the Borel $b$ subalgebras of $\hat g_0$ are
defined as:
$$\eqalign{
  h  &=\hat g^0\cap\hat g_0=\bbc (H_1-H_3)\oplus
  \bbc(H_2-H_4)\cr
  \eta &=\left(\mathop{\oplus}\limits_{k<0}\ \hat g^{k/2}\right)
  \cap\hat g_0=\mathop{\oplus}\limits_i\ \bbc\ E_{-i}
  \mathop{\oplus}\limits_{i<j}\ \bbc\ (E_{--})_{ij}\oplus\
  \bbc\ (E_{---})_{12}\cr
  b &=h\oplus\eta\ .\cr}
  \eqno(2.18)$$

For later use, we give hereafter the $4\times 4$ matrix representation
of the basic generators of $\hat g$.
$$\eqalign{
  E_{+1} &=\left(\matrix{ 0&0&1&0\cr 0&0&0&0\cr
  0&0&0&0\cr 0&0&0&0\cr}\right)\ ;\qquad
  E_{+2} =\left(\matrix{ 0&0&0&0\cr 0&0&0&0\cr
  0&1&0&0\cr 0&0&0&0\cr}\right)\cr
  &\cr
   E_{+3} &=\left(\matrix{ 0&0&0&0\cr 0&0&0&1\cr
  0&0&0&0\cr 0&0&0&0\cr}\right)\ ;\qquad
  E_{+4} =\left(\matrix{ 0&0&0&0\cr 0&0&0&0\cr
  0&0&0&0\cr \mu^2&0&0&0\cr}\right)\ .\cr}                  \eqno(2.19)$$
Note finally that the sum of these odd generators, which appear in the
$\hat A(1\vert 1)^{(1)}$ Lax formalism as we shall see later on, obeys:
$$\eqalign{
  \Lambda_+ &=\sum^4_{a=1}\ E_{+a}\cr
  (\Lambda_+)^4 &=\mu^2_{++}\ I_{4\times 4}\cr
  {\rm CSW} (\Lambda_+) &= {1\over 2}\ ,\cr}                   \eqno(2.20)$$
where $I_{4\times 4}$ is the $4\times 4$ identity matrix.

\SUBSECTION{The $\hat A(2\vert 2)^{(1)}$ affine Lie superalgebra}

This Lie superalgebra is involved in the construction of $N=2$ $W_3$
currents. As in the $n=1$ case, the $A(2\vert 2)$ superalgebra is
obtained by factorizing out the one dimensional centre of $SL(3\vert 3)$
namely, the $6\times 6$ unit matrix. $SL(3\vert 3)$, which has rank 5 and
dimension $d=17+18$, may be realized in terms of $6\times 6$ square
matrices with vanishing supertrace. The purely fermionic simple root
system $\{\alpha_a,a=(i,6)\}$ of the affine Lie superalgebra
$\widehat{SL}(3\vert 3)^{(1)}$ has a Cartan matrix
$$K_{ab}=\left(\matrix{
  0&1&0&0&0&1\cr
  -1&0&1&0&0&0\cr
  0&-1&0&1&0&0\cr
  0&0&-1&0&1&0\cr
  0&0&0&-1&0&1\cr
  -1&0&0&0&-1&0\cr}\right)\ .
\eqno(2.21)$$

The $(17+18)$ generators of $SL(3\vert 3)$ Lie superalgebra associated
with the purely fermionic simple root system $\{\alpha_i, i=1,\dots 5\}$
may be constructed out of the fundamental ones: $E_{\pm\alpha_i}$
as in Eqs.(2.4), (2.5). We shall not use this method since we can do better.
The point is that the basic generators $E_{\pm a}$ have conformal spin
weights equal to $\pm{1\over 2}$ and are realized in terms of $6\times 6$
supermatrices. This means that these $6\times 6$ matrices and fortiori
the $4\times 4$ previous ones Eqs.(2.19) carry the conformal spin
structure introduced earlier Eqs.(2.15)--(2.17). The same logic implies
that this feature is present at the level of the six dimensional $SL(3\vert 3)$
vector representation space $V_6$. To better understand this property,
let $\nu_a=\vert a>; a=(i,6)$ denote the canonical vector basis of $V_6$.
The $\nu_a$'s are column vectors with elements
$$(\nu_a)_b=\delta_{ab}\ .
\eqno(2.22)$$
The dual vector space $V^*_6$ is generated by row vectors
$(\nu^*_a)=<a\vert$. The hermitian bilinear form on $V_6$ is given by
$$\nu^*_a(\nu_b)=\ <a\vert b>\ =\delta_{ab}\ .                  \eqno(2.23)$$
The $SL(3\vert 3)\ \ 6\times 6$ matrix generators can be represented
in terms of the ket--bras vectors as $\vert a><b\vert$. Since the
$SL(3\vert 3)$ generators carry conformal spin weights, one concludes
that the kets $\vert a>$ and their duals $<a\vert$ carry themselves
conformal spin weights. The latter read in the $SL(3\vert 3)^{(1)}$
case as:
$$\vbox{\def\tablerule{\noalign{\hrule}}
\offinterlineskip\baselineskip=16pt\halign{\strut#&
\vrule#&
\hfil#\hfil&\vrule#&\hfil#\hfil&\vrule#&\hfil#\hfil
                &\vrule#&\hfil#\hfil&\vrule#&
\hfil#\hfil&\vrule#&\hfil#\hfil&\vrule#&\hfil#\hfil
                &\vrule#&\hfil#\hfil&\vrule#\cr

&\omit \quad Vectors\quad&&\quad $\nu_1$\quad&&\quad $\nu_2$\quad
&&\quad $\nu_3$\quad&&\quad $\nu_4$\quad&&\quad $\nu_5$\quad
&&\quad $\nu_6$\quad&\omit\cr
\tablerule
&\omit \quad CSW&&\quad ${5\over 2}$&&\quad ${3\over 2}$
&&\quad ${1\over 2}$&&\quad 2&&\quad 1&&\quad 0
&\omit\cr}}\ .
       \eqno(2.24)$$
As required by the hermitian bilinear form Eq.(2.23), the CSW of the dual
vectors $\nu^*_a$ have the opposite values of the CSW of $\nu_a$'s.
In the next section we shall give the generic formula of the CSW of the
canonical vectors $\nu_a$ of the $\hat A(n-1\vert n-1)^{(1)}$ vector
representation space $V_{2n}$.

Using this result, the basic generators $E_{\pm i}$ of the $SL(3\vert 3)$
Lie superalgebra associated with the purely fermionic simple root
system $\{\alpha_i, i=1,\dots 5\}$ read as:
$$\vbox{\def\tablerule{\noalign{\hrule}}
\offinterlineskip\baselineskip=16pt\halign{\strut#&
\vrule#&
\hfil#\hfil&\vrule#&\hfil#\hfil&\vrule#&\hfil#\hfil&\vrule#\cr
&&\quad Basic generators: $E_{+i}$\quad
&&\quad CSW $(E_{+i})={1\over 2}$\quad&\cr
\tablerule
&&$E_{+1}=\vert 1><4\vert$&&${5\over 2}-2$&\cr
&&$E_{+2}=\vert 4><2\vert$&&$2-{3\over 2}$&\cr
&&$E_{+3}=\vert 2><5\vert$&&${3\over 2}-2$&\cr
&&$E_{+4}=\vert 5><3\vert$&&$1-{1\over 2}$&\cr
&&$E_{+5}=\vert 3><6\vert$&&${1\over 2}-0$&\cr
\tablerule}}
     \eqno(2.25)$$
The $E_{-i}$ generators are given by the adjoint of the $E'_{+i}$'s, e.g.
$E_{-1}=(E_{+1})^+=\vert 4><1\vert$ and so on. The generators $E_{+6}$
associated with the highest root $\alpha_6$ are given by
$$\eqalign{
  (E_{+6}) &=\mu^3_{++}\vert 6><1\vert\cr
  (E_{-6}) &=\mu^{-3}_{++}\vert 1><6\vert\cr}
\eqno(2.26)$$
Their CSW's are respectively ${1\over 2}$ and $-{1\over 2}$ as may be read
from Eqs.(2.24) and (2.15). Note that the fermionic generators $E_{+i}$
Eqs.(2.25) satisfy the remarkable property
$$E_{+i}\ E_{+j}=\delta_{j,i+1}\ E_{+i}\ E_{+i+1}
\eqno(2.27)$$
thanks to the hermitian bilinear form $<\vert >$ of $V_6$ Eq.(2.23). The
remaining $(5+0)+2(6+4)$ generators of $SL(3\vert 3)$ are obtained
by taking the supercommutators of the $E_{\pm_i}$'s and their
combinations. These calculations turn out to be greatly simplified, once
more, because of the properties of the hermitian form. We find:
$$\eqalign{
  H_1 &=\vert 1><1\vert +\vert 4><4\vert\cr
  H_2 &=\vert 4><4\vert +\vert 2><2\vert\cr
  H_3 &=\vert 2><2\vert +\vert 5><5\vert\cr
  H_4 &=\vert 5><5\vert +\vert 3><3\vert\cr
  H_5 &=\vert 3><3\vert +\vert 6><6\vert\cr}                     \eqno(2.28)$$
and by help of Eq.(2.27)
$$\eqalign{
  (E_{++})_{ij} &= \delta_{j,i+1}\ E_{+i}\ E_{+j}\cr
  (E_{+++})_{ijk} &=\delta_{k, j+1}\ \delta_{j,i+1}\ E_{+i}\ E_{+j}\ E_{+k}\cr
  (E_{++++})_{ijk\ell} &=\delta_{\ell ,k+1}\ \delta_{k,j+1}\ \delta_{j,i+1}\
  E_{+i}\ E_{+j}\ E_{+k}\ E_{+\ell}\cr
  (E_{+++++})_{12345} &= E_{+1}\cdot E_{+2}\cdot E_{+3}\cdot
  E_{+4}\cdot E_{+5}\ .\cr}
\eqno(2.29)$$
the $(E_{-r})_{i_1\dots , i_r}$'s, $i_1<i_2<\dots <i_r;1\leq r\leq 6$ are
given by the adjoints of Eqs.(2.29). These relations, which seem cumbersome,
have in fact a very simple form, e.g.
$$\eqalign{
  (E_{++})_{12} &=\vert 1><2\vert\cr
  (E_{+++++})_{12345} &= \vert 1><6\vert\ .\cr}
\eqno(2.30)$$
The CSW of these generators which is read from the tableau (2.24) is
respectively given by ${5\over 2}-{3\over 2}=1$ and ${5\over 2}-0=
{5\over 2}$. The charges carried by the generators Eqs.(2.29) are then
related to the conformal spin weight.

We end this subsection by introducting the matrix operators $\Lambda_+$,
the analogue of Eqs.(2.20). It is given by the sum of the conformal spin
${1\over 2}$ generators, namely
$$\Lambda_+ =\sum^6_{a=1}\ E_{+a}\ .
\eqno(2.31)$$
This is also a conformal spin ${1\over 2}$ object exhibiting the following
properties:

\noindent 1)\ Decomposability:
$$\Lambda_+ =\Lambda^+_++\Lambda^-_+                          \eqno(2.32)$$
where
$$\eqalign{
  \Lambda^+_+ &=\vert 1><4\vert +\vert 2><5\vert +
  \vert 3><6\vert\cr
  \Lambda^-_+ &=\vert 4><2\vert +\vert 5><3\vert +\mu^3
  \vert 6><1\vert\ .\cr}
    \eqno(2.33)$$
The $\pm$ upper charges carried by $\Lambda^\pm_+$ should not be
confused with the space--time lower charge \ +.

\noindent 2) \ Nilpotency:
$$\eqalign{
  \{\Lambda^+_+,\Lambda^+_+\} &=2(\Lambda^+_+)^2=0\cr
  \{\Lambda^-_+,\Lambda^-_+\} &=2(\Lambda^-_+)^2=0\ .\cr}  \eqno(2.34)$$
Taking the supercommutators of $\Lambda^+_+$ and $\Lambda^-_+$,
one obtains the third property, that is, reducibility:
$$\eqalign{
  \Lambda_{++} &=\{\Lambda^+_+,\Lambda^-_+\} =\lambda_{++}
  \oplus\lambda_{++}\cr
  \Sigma_{++} &=[\Lambda^+_+,\Lambda^-_+] =\lambda_{++}
  \ominus\lambda_{++}\ ,\cr}
\eqno(2.35)$$
where $\lambda_{++}$ is a $3\times 3$ matrix given by
$$\lambda_{++} =\left(\matrix{ 0&1&0\cr 0&0&1\cr
  \mu^3&0&0\cr}\right)\ .
\eqno(2.36)$$
Note that Eqs.(2.35) leave invariant the two three--dimensional subspaces
of $V_6$ respectively generated by $\{\vert k>\}$ and $\{\vert k+3>\},
1\leq k\leq 3$. This feature will be used when we consider the manifestly
$N=2\ U(1)$ Lax formalism, see Subsection 4.3. The last property is
periodicity since
$$(\Lambda_+)^6=\mu^3_{++}\ I_{6\times 6}\ ,
\eqno(2.37)$$
or equivalently
$$(\Lambda_{++})^3=(\Sigma_{++})^3=\mu^3_{++}\
  I_{6\times 6}\ ,
     \eqno(2.38)$$
where $I_{6\times 6}$ is the $6\times 6$ unit matrix.

\SECTION{MORE ON THE $\hat A(n-1\vert n-1)^{(1)}$ SUPERALGEBRA}

Here we generalize the previous results to any positive integer $n>1$.
The $SL(n\vert n)$ vector representation space $V_{2n}$ is a graded
space generated by the complete orthogonal set $\{\vert k>,\vert k+n>~;$
$1\leq k\leq n\}$. The conformal spin weights of these basis vectors
generalizing Eqs.(2.24) are given by
$$\vbox{\def\tablerule{\noalign{\hrule}}
\offinterlineskip\baselineskip=12pt\halign{\strut#&
\vrule#&
\hfil#\hfil&\vrule#&\hfil#\hfil&\vrule#&\hfil#\hfil
                &\vrule#&\hfil#\hfil&\vrule#&
\hfil#\hfil&\vrule#&\hfil#\hfil&vrule#\cr

&&\quad Vectors\quad&&$\vert k>$
&&\quad $\vert k+n>$\quad
&&$<k\vert$&&\quad $<k+n\vert$\quad&\cr
\tablerule
&&CSW&&\quad $n-k+{1\over 2}$\quad&&$n-k$
&&\quad $k-n-{1\over 2}$\quad&&\quad $k-n$&\cr
\tablerule}}\ .

         \eqno(3.1)$$
 From this table, one sees that the $n$ first vectors $\vert k>$ of $V_{2n}$
behave as conformal spinors whereas the remaining $n\vert k+n>$ ones
behave as conformal bosons. Conformal odd operators are then
represented by matrices made of an odd ket and an even bras or
inversely. Taking into account this feature and using Eq.(3.1), one finds
that the $2(2n-1)$ basic generators $E_{\pm i}$ associated with purely
fermionic simple root system $\{\alpha_i,i=1,\dots 2n-1\}$ of
$SL(n\vert n)$ read as
$$\eqalign{
  E_{+(2k-1)} &=\vert k><k+n\vert\qquad\quad\ \ 1\leq k\leq n\cr
  E_{+2k} &=\vert k+n><k+1\vert\qquad 1\leq k\leq n-1\cr}   \eqno(3.2)$$
and
$$\eqalign{
  E_{-(2k-1)} &=\vert k+n><k\vert\qquad\quad\ \ \ 1\leq k\leq n\cr
  E_{-2k} &=\vert k+1><k+1\vert\qquad 1\leq k\leq m-1\ .\cr}  \eqno(3.3)$$
It is interesting to check, with the help of Eq.(3.1), that the above $2n+2n$
matrices have respectively CSW's ${1\over 2},{1\over 2},-{1\over 2}$
and $-{1\over 2}$. Here we also have the analogue of Eqs.(2.29) namely:
$$\eqalign{
  E_{+(2k-1)}\cdot E_{+2\ell} &=\delta_{k\ell}\ \vert k><k+1\vert\cr
  E_{+(2k-1)}\cdot E_{+(2\ell -1)} &= 0\cr
  E_{+(2k-1)}\cdot E_{+(2k-1)} &=\delta_{k\ \ell+1}\
  \vert k+n-1><k+n\vert\cr
  E_{+2\ell}\cdot E_{+2k} &= 0\ ,\cr}
\eqno(3.4)$$
thanks to the $V_{2n}$ hermitian form $<\vert >$. The remaining
$(2n^2-1)+2(n^2-2n+1)$ are obtained by help of Eqs.(3.4). The $(2n-1)\
SL(n\vert n)$ Cartan generators read as:
$$\eqalign{
  H_{2k-1} &=\vert k><k\vert +\vert k+n><k+n\vert\qquad\qquad\quad \
  1\leq k\leq n\cr
  H_{2k} &= \vert k+1><k+1\vert +\vert k+n><k+n\vert\qquad
  1\leq k\leq n-1\ ,\cr}
       \eqno(3.5)$$
and the $2n(n-1)+2(n^2+1-2n)$ higher CSW ones are collected in the
following generic formula:
$$\eqalign{
  (E_{+r})_{k_jk_{j+1}\dots k_{j+2}} &=\mathop{\Pi}\limits^{j+r}_{s=j}\
  (E_{k_s})\cr
  (E_{-r})_{k_jk_{j+1}\dots k_{j+2}} &= (E_{+r})^+_{k_j, k_{j+1}
  \dots k_{j+2}}\ ,\cr}
         \eqno(3.6)$$
where $1\leq j<j+r\leq 2n-1$. These relations depend on two indices $j$
and $r$ and their CSW is equal to ${r\over 2}$ and $-{r\over 2}$
respectively. Moreover, the conformal spin $\pm{1\over 2}$ generators
associated with the $SL(n\vert n)$ highest root $\alpha_{2n}$
generalizing Eqs.(2.26) are given by
$$\eqalign{
  E_{+2n} &=\mu^n\vert 2n><1\vert\cr
  E_{-2n} &=\mu^{-n}\vert 1><2n\vert\cr}
\eqno(3.7)$$
and the corresponding Cartan matrix generator $H_{2n}$ reads as
$$H_{2n}=\vert 1><1\vert +\vert 2n><2n\vert\ .
\eqno(3.8)$$
Note by the way that the $SL(n\vert n)$ centre is given by
$$\sum^n_{k=1}\ H_{2k-1}=\sum^n_{k=1}\ H_{2k} =
  I_{2n\times 2n}\ .
       \eqno(3.9)$$

The properties of the sum of the conformal spin ${1\over 2}\
\widehat{SL}(3\vert 3)^{(1)}$ generators Eqs.(2.31)--(2.38) are also
present in the general $\widehat{SL}(n\vert n)^{(1)}$ case. Indeed we
have:

\noindent 1)\ Decomposability:
$$\eqalign{
  \Lambda_+ &=\Lambda^+_+ +\Lambda^-_+\cr
  \Lambda^+_+ &=\sum^n_{k=1}\ E_{2k-1}\cr
  \Lambda^-_+ &=\sum^n_{k=1}\ E_{+2k}\ .\cr}
\eqno(3.10)$$
2)\ Nilpotency:
$$\eqalign{
  (\Lambda^+_+)^2 &= 0\cr
  (\Lambda^-_+)^2 &= 0\ .\cr}
\eqno(3.11)$$
3)\ Reducibility:
$$\eqalign{
  \Lambda_{++} &=\{\Lambda^+_+,\Lambda^-_+\} =\lambda^{(n)}_{++}
  \oplus\lambda^{(2n)}_{++}\cr
  \Sigma_{++} &=[\Lambda^+_+,\Lambda^-_+]=\lambda^{(n)}_{++}
  \ominus\lambda^{(2n)}_{++}\ ,\cr}
\eqno(3.12)$$
where $\lambda^{(n)}_{++}$ and $\lambda^{(2n)}_{++}$ are $n\times n$
matrices respectively given by
$$\eqalignno{
  \lambda^{(n)}_{++} &=\sum^{n-1}_{k=1}\ \vert k><k+1\vert +
  \mu^3\vert n><1\vert &(3.13)\cr
  \lambda^{(2n)}_{++} &=\sum^{n-1}_{k=1}\ \vert k+n><k+n+1\vert +
  \mu^n\vert 2n><n+1\vert\ .           &(3.14)\cr}$$
Eqs. (3.13) and (3.14) are nothing but the sum of the generators
associated with the simple roots of the $\widehat{SL}(n)^{(1)}$ affine
Lie algebra. These quantities, which in our language are conformal spin one
objects, play a crucial role in the generalized $\widehat{SL}(n)^{(1)}$
hierarchy [3, 5].

\noindent 4) Periodicity:
$$(\Lambda_+)^{2n}=(\Lambda_{++})^n=(\Sigma_{++})^n=
  \mu^n_{++}\ I_{2n\times 2n}\ ,
\eqno(3.15)$$
where $I_{2n\times 2n}$ is the $2n\times 2n$ unit matrix.

For later use, let us illustrate the above relations in the $n=2$ case.
This will also be an opportunity to compare our method based on the
conformal spin gradation of $\hat g$ with the standard analysis of
subsection 2.1. The CSW of the canonical basis vectors of the $SL(2\vert 2)$
representation vector space $V_4$ read as
$$\vbox{\def\tablerule{\noalign{\hrule}}
\offinterlineskip\baselineskip=15pt\halign{\strut#&
\vrule#&
\hfil#\hfil&\vrule#&\hfil#\hfil&\vrule#&\hfil#\hfil
                &\vrule#&\hfil#\hfil&\vrule#&
\hfil#\hfil&\vrule#&\hfil#\hfil&vrule#\cr

&\omit&&&\quad $\vert 1>$\quad
&&\quad $\vert 2>$\quad
&&\quad $\vert 3>$\quad &&\quad $\vert 4>$\quad&\cr
\tablerule
&\omit&\quad CSW\quad&&${3\over 2}$&&${1\over 2}$
&&1&&0&\cr
\tablerule}}\ .
        \eqno(3.16)$$
The fundamental $SL(2\vert 2)$ conformal spin ${1\over 2}$ generators
are given by
$$\eqalign{
  E_{+1} &=\vert 1><3\vert\cr
  E_{+2} &=\vert 3><2\vert\cr
  E_{+3} &=\vert 2><4\vert\ .\cr}
\eqno(3.17)$$
These relations together with
$$E_{+4} =\mu^2_{++}\vert 4><1\vert\ ,
\eqno(3.18)$$
coincide exactly with Eqs.(2.19). The spin $-{1\over 2}$ generators are
obtained by taking the adjoint of these equations. The Cartan generators,
which have vanishing CSW, read as
$$\eqalign{
  H_1 &=\vert 1><1\vert +\vert 3><3\vert\cr
  H_2 &=\vert 2><2\vert +\vert 3><3\vert\cr
  H_3 &=\vert 2><2\vert +\vert 4><4\vert\cr
  H_4 &=\vert 1><1\vert +\vert 4><4\vert\ .\cr}                \eqno(3.19)$$
The conformal spin 1 and ${3\over 2}$ generators involving the
generators (3.17) are immediately obtained with the help of Eqs.(3.4)
$$\eqalign{
  (E_{++})_{12} &= \vert 1><2\vert\cr
  (E_{++})_{23} &= \vert 3><4\vert\cr
  (E_{+++})_{123} &=\vert 1><4\vert\ .\cr}
\eqno(3.20)$$
Adjoints of these relations give the conformal spin $-1$ and $-{3\over 2}$
generators. The $\Lambda^\pm_+$ as well as $\Lambda_{++}$ and
$\Sigma_{++}$ read as
$$\eqalignno{
  \Lambda^+_+ &= \vert 1><3\vert + \vert 2><4\vert\cr
  \Lambda^-_+ &= \vert 3><2\vert + \mu^2\vert 4><1\vert\ . &(3.21)\cr
  &\cr
  \Lambda_{++} &= \lambda_{++}\oplus\lambda_{++}\cr
  \Sigma_{++} &=\lambda_{++}\ominus\lambda_{++}\ ,           &(3.22)\cr}$$
where $\lambda_{++}$ is the sum of the $\widehat{SL}(2)^{(1)}$
conformal spin one generators
$$\lambda_{++} =\left(\matrix{ 0&1\cr \mu^2 &0\cr}\right)\ . \eqno(3.23)$$

\SECTION{THE $\hat A(n-1\vert n-1)^{(1)}$ MIURA TRANSFORMATION}

Here we use the conformal spin gradation of $\hat A(n-1\vert n-1)^{(1)}$
to study the $N=2$ supersymmetric Miura transformation leading to the
general $N=2$ superfield realizations of $N=2$ higher conformal spin
conserved currents. The special example of $N=2$ $W_3$ symmetry as well
as comparison with recent results will be considered in the forthcoming
sections. Our aim here is as follows. First, set up the fundamentals of
a manifestly $N=2$ Miura transformation which, from our point of view,
is the right way to deal with $N=2$ higher conformal spin symmetries. To
our knowledge this programme has not been studied in the literature.
Partial results will be given here. Second, we develop a new method, based
on $U(1)$ charge decompositions and allowing the obtainment of the
$N=2$ superfield realizations of the $W$ currents from the $N=1$ language
of $\hat A(n-1\vert n-1)^{(1)}$.

\SUBSECTION{Generalities on $W\widehat{SL}(n)^{(1)}$ Miura
transformation}

We start by considering the bosonic generalized $\widehat{SL}(n)^{(1)}$
hierarchies. There, the $N=0$ Miura transformation may be obtained as
follows: First, one considers the first order differential operator $\ell$
valued in the $\widehat{SL}(n)$ affine algebra [3, 5]
$$\ell =\partial /\partial_x +q(t,x) +\lambda
\eqno(4.1)$$
where $q(t,x)$ is a two dimensional hermitian field valued in the
$\widehat{SL}(n)$ Borel subalgebra. $\lambda$ is the sum of the basic
chevalley generators corresponding to the $\widehat{SL}(n)$ positive
simple roots. In terms of the canonical basis $\{ \vert k>, 1\leq k\leq n\}$
of the $SL(n)$ representation vector space $V_n$, the $n\times n$\ \
$\lambda$ matrix reads as
$$\lambda =\sum^{n-1}_{k=1}\ \vert k><k+1\vert +
  \nu\vert n><1\vert\ ,
   \eqno(4.2)$$
where $\nu$ is a spectral parameter. For later use, let us rewrite
Eq.(4.1) as
$$\ell_{++} =\partial_{++} +q_{++} +\lambda_{++}\ ,
\eqno(4.3)$$
where
$$\eqalign{
  \partial_{++} &= \partial /\partial_{x_{--}}\sim\partial_z\cr
  q_{++} &= q_{++}(x_{--},x_{++})\sim q_z(z,\bar z)\ . \cr}        \eqno(4.4)$$
To make contact with the usual analysis Eq.(4.1), one should interpret
$x_{++}$ as the time coordinate. Thus, in the new language, one sees that
$\ell_{++}$ behave as a conformal spin one operator. The second step is to
make two different gauge choices of the $\widehat{SL}(n)^{(1)}$ valued
field $q_{++}$. The first one is given by the so--called diagonal gauge
in which Eq.(4.3) takes the form
$$\ell^0_{++} =\partial_{++} +\sum^n_{i=1}\ (q_{++})_i\
  \vert i><i\vert +\lambda_{++}\ ,
 \eqno(4.5)$$
where the $(q_{++})_i$'s obey the $SL(n)$ traceless condition:
$$\sum^n_{i=1}\ (q_{++})_i = 0\ .
\eqno(4.6)$$
The second gauge choice is given by the canonical gauge in which the
$(n-1)$ physical degrees of freedom of $q_{++}$ are distributed as:
$$\ell^{(1)}_{++} =\partial_{++} +\sum^n_{i=2}\ u_i\vert i><1\vert +
  \lambda_{++}\ ,
         \eqno(4.7)$$
where the $u'_k$'s behave as conformal spin $k$ objects. The traceless
condition implies $u_1=0$.\hfil\break
The final step is to look for a polynom in $\ell_{++}$; $P(\ell_{++})$, such
that
$$<n\vert P(\ell_{++})\vert n>\ =\nu^n\ .
\eqno(4.8)$$
Moreover, as the r.h.s. of the above relation is invariant under the gauge
transformations generated by the exponential of $\widehat{SL}(n)^{(1)}$
nilpotent matrices [3, 5], one concludes
$$<n\vert P(\ell^{(0)}_{++})\vert n>\ =\ <n\vert P(\ell^{(1)}_{++})\vert n>\ .

               \eqno(4.9)$$
Straightforward algebraic analysis leads to the $\widehat{SL}(n)^{(1)}$
Miura transformation, namely
$$\partial^n_{++} -\sum^n_2\ u_k\ \partial^{n-k}_{++} =
  \mathop{\Pi}\limits^n_{j=1}\ (\partial_{++} -q_{++})_j       \eqno(4.10)$$
where we have used the convention notation:
$$(\partial_{++} -q_{++})_j=\partial_{++} -q_{++j}\ .
\eqno(4.11)$$
Expanding the r.h.s. of Eq.(4.10), one gets the field realization of the
conformal spin $k$ conserved currents. Note finally that for a given
value of $n$, one has $(n-1)$ conformal currents with conformal spin
$k$ with $2\leq k\leq n$. A similar structure will be found in the
$\hat A(n-1\vert n-1)^{(1)}$ case.

\SUBSECTION{$W\hat A(n-1\vert n-1)^{(1)}$ Miura transformation}

Recall that supersymmetric extension of bosonic integrable theories is
possible only for simple and affine Lie superalgebras which possess a
purely fermionic simple root system. These superalgebras are given by
[19, 8, 14]
$$SL(n\pm 1\vert n);\ \  Osp(m\vert 2n),\  (m=2n, 2n+2, 2n\pm 1):\ \
D(2\vert 1;\alpha )\ ;
  \eqno(4.12)$$
for the finite dimensional case and
$$\eqalign{
  \widehat{SL}(n\vert n)^{(1)};\ \ \widehat{Osp}(2n+2\vert 2n)^{(1)};\ \
  \hat D(2\vert 1:\alpha )^{(1)}\cr
  \widehat{SQ}(2n+1)^{(2)};\ \ \widehat{SL}(n\vert n)^{(2)};\ \
  \widehat{Osp}(2n\vert 2n)^{(2)}\cr}
\eqno(4.13)$$
in the infinite dimensional case.

In the $N=1$ supersymmetric case, one introduces the superoperator
generalizing Eq.(4.1)
$${\cal D}_+=D_++Q_++\Lambda_+\ ,
\eqno(4.14)$$
where $D_+$ is the usual $N=1$ covariant derivatives satisfying
$D^2_+=\partial_{++}$. $Q_+$ is a hermitian $N=1$ superfield valued in the
Borel subalgebra of a given Lie superalgebra of Eqs.(4.12) and (4.13).
Here we shall focus our attention on $\hat A(n-1\vert n-1)^{(1)}\subset
\widehat{SL}(n)^{(1)}$. $\Lambda_+$ is the sum of the conformal spin
${1\over 2}$ generators given by Eqs.(3.10). The supersymmetric partner
of ${\cal D}_+$ is obtained by solving the following $N=1$ superalgebraic
equation
$$\eqalign{
  2{\cal L}_{++} &=\{{\cal D}_+,{\cal D}_+\}\cr
  0 &=[{\cal L}_{++},{\cal D}_+]\ .\cr}
\eqno(4.15)$$
We find
$${\cal L}_{++} =\partial_{++}+q_{++}+\Lambda_{++}\ ,
\eqno(4.16)$$
where $\Lambda_{++}$ is the sum of the conformal spin one generators
of the $\widehat{SL}(n)^{(1)}\oplus\widehat{SL}(n)^{(1)}$ affine algebra
given by Eqs.(3.12)--(3.14) and where
$$q_{++}=D_+Q_+ +(Q_+)^2+\{ Q_+,\Lambda_+\}\ .                  \eqno(4.17)$$
Note that we also have
$$[{\cal D}_+,{\cal D}_+]=0\ .
 \eqno(4.18)$$

The following step is to make two gauge choices of the $\widehat{SL}
(n\vert n)^{(1)}$ Lax operator ${\cal D}_+$ [5, 15]. The first one is given by
the diagonal gauge in which ${\cal D}_+$ takes the form
$${\cal D}^{(0)}_+=D_++\sum^{2n}_{i=1}\ (Q_+)_{ii}\vert i><i\vert +
  \Lambda_+
     \eqno(4.19)$$
where the $(Q_+)_{ii}$'s are subject to the $SL(n\vert n)$ supertraceless
condition, namely:
$$\sum^n_{k=1}\ \left(\xi_{+k}-\xi_{+(k+n)}\right) =0\ ,       \eqno(4.20)$$
with $\xi_{+k}=(Q_+)_{kk}$.

The second gauge choice is given by the following canonical form of
${\cal D}_+$:
$${\cal D}^{(1)}_+ =D_++\sum^n_{k=1}\left( U_{2k}\vert k+n><1\vert +
  U_{2k-1}\vert k><1\vert\right) +\Lambda_+\ ,                   \eqno(4.21)$$
where $U_{2k}$ and $U_{(2k-1)}$ behave as conformal spin $k$ and
$\left( k-{1\over 2}\right)$ objects respectively. Note that the
$SL(n\vert n)$ supertrace condition requires the vanishing of the
conformal spin ${1\over 2}$ current; $U_1=0$. Therefore, given a
positive integer $n$, the conformal spin $s$ of the $U$'s is bounded as
$$1\leq s\leq n,\quad 2s\in\bbn\ .
\eqno(4.22)$$
However, as we are interested in the $N=2$ Miura transformation based
on the affine Lie superalgebra $\hat A(n-1\vert n-1)^{(1)}\subset
\widehat{SL}(n\vert n)^{(1)}$, we should impose an extra condition on
the $Q_+$ degrees of freedom. Thus setting $\xi_{+2n}=0$ or equivalently
$U_{2n}=0$, Eq.(4.22) is modified as
$$1\leq s\leq n-{1\over 2};\quad n\geq 2\ .
\eqno(4.23)$$
Eq.(4.23) shows that for a given integer $n$, one has $(2n-2), N=1$
hermitian superfields describing $(n-1)$\ $N=2$ multiplets of integer
conformal spins $k$ with $1\leq k\leq n-1$. These $N=2$ multiplets
are constructed from the $U$'s of Eq.(4.21) as
$$J_k=U_{2k}\oplus U_{2k+1};\quad 1\leq k\leq n-1\ .            \eqno(4.24)$$
In terms of the $N=0$ conformal fields, the above relation reads as
$$J_k\sim\left( k, k+{1\over 2}, k+{1\over 2}, k+1\right)\ .    \eqno(4.25)$$

To obtain the $\hat A(n-1\vert n-1)$ Miura transformation in the $N=1$
superfield language, one should look for a polynom in the $N=1$ Lax
operator $P({\cal D}_+)$ obeying the equation
$$<2n\vert P({\cal D}_+)\vert 2n>\ = \mu^n_{++}\ .
\eqno(4.26)$$

As the r.h.s. of this relation is invariant under gauge transformations
generated by the exponential of $\hat A(n-1\vert n-1)^{(1)}$ nilpotent
matrices, we get
$$<2n\vert P({\cal D}^{(1)}_+)\vert 2n>\ =\ <2n\vert
  P({\cal D}^{(0)}_+)\vert 2n>\ ,
 \eqno(4.27)$$
or equivalently
$$D^{2n}_++\sum^{2n}_{i=2}\ U_i\ D^{2n-i}_+ =
  \mathop{\Pi}\limits^n_{j=1}\ (D_+-\xi_+)_j(D_+-\xi_+)_{j+n}  \eqno(4.28)$$
with
$$(D_+-\xi_+)_j=D_+-\xi_{+j}\ .
\eqno(4.29)$$
Eq.(4.28) defines the $\widehat{SL}(n\vert n)^{(1)}$ Miura
transformation. Solving Eqs.(4.20) and (4.23) in terms of $2(n-1)$
degrees of freedom as
$$\eqalign{
  (\xi_+)_{k+1} &=\psi_{+2k}+\psi_{+(2k+1)}\cr
  (\xi_+)_{k+n+1} &=\psi_{+(2k+1)}+\psi_{+(2k+2)};\quad
  1\leq k\leq n-1  \cr}
    \eqno(4.30)$$
with
$$\eqalign{
  \psi_{+(-1)} &= \psi_{+0}=0\cr
  \psi_{+(2n-1)} &=\psi_{+2n}=0\ ,\cr}
\eqno(4.31)$$
Eq.(4.28) can be rewritten as
$$D^{2n}_++\sum^{2n-1}_{i=2}\ U_i\ D^{2n-i}_+=
  \mathop{\Pi}\limits^{n-1}_{k=0}\ \left( D_+-\psi_{+2k}-\psi_{+2k+1}
  \right)\left( D_+-\psi_{+2k+1}-\psi_{+2k+2}\right)\ .          \eqno(4.32)$$
Note that the r.h.s. of Eqs.(4.28) and (4.32) is built out of blocks of two
elements. This is a property of the $W\ \widehat{SL}(n\vert n)^{(1)}$
Miura transformation.

\SUBSECTION{The $N=2\ U(1)$ Lax formalism}

To our knowledge, no manifestly $N=2$ super Miura transformation has
been written down in the literature. Known formulations [15, 16] are
based on the breaking of $N=2$ supersymmetry down to $N=1$. Here
we give the a.b.c. of this analysis. Progress will be reported elsewhere [21].

We begin by noting that a manifestly $N=2$ supersymmetric Lax formalism
should keep track with the $U(1)$ Kac--Moody charge of the $N=2$
superconformal algebra [12, 13]. The point is that in $N=2$ CFT's, the $U(1)$
charge plays a crucial role as it is related to the conformal weight $h$
[12, 13]. The so--called chiral rings [13], the $N=2$ Landau Ginzburg
theories [22] and the $N=2$ (anti) topological theories [23] are all
based on this charge. Suppose then the existence of two $U(1)$ charge
conjugate Lax type superoperators, generalizing Eqs,(4.3) and (4.14),
given by
$$\eqalign{
  {\cal D}^+_+ &= D^+_++Q^+_++G^+_+\cr
  {\cal D}^-_+ &= D^-_++Q^-_++G^-_+\ ,\cr}
\eqno(4.33)$$
where $D^\pm_+$ are the usual $N=2$ covariant derivatives obeying the
$N=2\ U(1)$ superalgebra
$$\eqalign{
  \{ D^+_+,D^-_+\} &= 2\partial_{++} \qquad\qquad\qquad\quad {\rm (a)}\cr
  \{ D^+_+,D^+_+\} &=\{D^-_+,D^-_+\}=0\ .\quad\quad\ \  {\rm (b)}\cr}

             \eqno(4.34)$$
As a convention notation, the $\pm$ upper indices refers to the $U(1)$
charges. The $Q^+_+$ and $Q^-_+$ are $N=2$ superfields valued in the
$\widehat{SL}(n\vert n)^{(1)}$ Lie superalgebra and $G^+_+$ and $G^-_+$
are conformal spin ${1\over 2}$ operators carrying $+1$ and $-1$
$U(1)$ charges respectively. At this level $Q^\pm_+$ and $G^\pm_+$ are
unknown quantities that should be specified by solving some constraint
equations. Note that the $\widehat{SL}(n\vert n)^{(1)}$ supertraceless
condition together with the factorization of the centre $c$ show that
$Q^\pm_+$ carry $2(n-1)\ N=2$ superfield degrees of freedom, i.e. twice
the desired number involved by the $W\hat A (n-1\vert n-1)$ Miura
transformation as shown by Eqs.(4.24) and (4.25).

Our constraints are of two types: The first set is given by
$$\eqalign{
  \{{\cal D}^+_+,{\cal D}^+_+\} &=0\qquad\qquad\qquad {\rm (a)}\cr
  \{{\cal D}^-_+,{\cal D}^-_+\} &=0\ .\qquad\qquad\quad \ \ {\rm (b)}\cr}

               \eqno(4.35)$$
The second set of constraints reads as
$$[{\cal D}^+_+,{\cal D}^-_+] ={\cal Z}^{[+-]}_{++}
\eqno(4.36)$$
where ${\cal Z}^{[+-]}_{++}$ is pseudo real conformal spin one operator valued
in $[\widehat{gL}(n)^{(1)}/gL(1)^{(1)}]\ominus
[\widehat{gL}(n)^{(1)}/$\hfil\break
$gL(1)^{(1)}]$
affine Lie algebra. It has a vanishing trace.

The first set of constraints is natural since it generalizes the identities
(4.34b). It implies
$$\eqalignno{
  D^+_+Q^+_+ &= 0 &(4.37)\cr
  D^+_+G^+_+ &= 0 &(4.38)\cr
  (Q^+_+)^2 &= 0 &(4.39)\cr
  (G^+_+)^2 &= 0 &(4.40)\cr
  \{ Q^+_+,G^+_+\} &= 0\ .    &(4.41)\cr}$$
together with their $U(1)$ charge conjugates. Let us comment briefly
these equalities. Eqs.(4.37) and (4.38) tell us that $Q^+_+$ and $G^+_+$
are $N=2$ chiral superfields. $Q^-_+$ and $G^-_+$ are then
antichiral superfields. However, as in the $N=0$ type I hierarchy
Eq.(2.13), we take $G^\pm_+$ to be superspace coordinates independent.

Eqs.(4.39)--(4.41) are Lie superalgebraic relations. All of them do not
reduce the $Q^\pm_+\ N=2$ superfield degrees of freedom. Indeed,
Eq.(4.39) is identically satisfied in the diagonal gauge $Q^{+(0)}_+$.
Eq.(4.40) is a nilpotency property and can be solved by $\Lambda^+_+$
Eq.(3.13). The last relation is a consistency condition.

The second set of constraints is less natural. Putting Eqs.(4.33) back into
Eq.(4.36), we get
$${\cal Z}^{[+-]}_{++} =\nabla_{++} +J^{[+-]}_{++}
  +\Sigma^{[+-]}_{++}
  \eqno(4.42)$$
where
$$\eqalignno{
  \nabla_{++} &= [D^+_+,D^-_+] &(4.43)\cr
  J^{[+-]}_{++} &= D^+_+Q^-_+-D^-_+Q^+_++[Q^+_+,Q^-_+] &(4.44)\cr
  \Sigma^{[+-]}_{++} &= [\Lambda^+_+,\Lambda^-_+]  &(4.45)\cr}$$
together with
$$[Q^+_+,G^-_+] =0
          \eqno(4.46)$$
and its $U(1)$ conjugate. Let us comment on these relations.
$\nabla_{++}$ is a conformal spin one operator behaving as $\partial_{++}$.
It has no $N=1$ analogue as shown by Eq.(4.18). $J^{[+-]}_{++}$ is a
composite object and has a conformal weight equal to one. It should be
compared with the $N=2\ U(1)$ current [24, 25, 12]. Eq.(4.45) is already
encountered in Section 3. See Eqs.(2.35, 2.36), (3.12--3.14) and (3.22, 3.23).

Eqs.(4.46) are necessary conditions in order that the bosonic operator
${\cal Z}^{[+-]}_{++}$ to be valued in the
$[\widehat{gL}(n)^{(1)}/gL(1)^{(1)}]\ominus
[\widehat{gL}(n)^{(1)}/gL(1)^{(1)}]$
algebra. We shall not discuss the Lie superalgebraic solutions of these
equations. We shall suppose, however, that they admit a solution.

The vanishing of the trace of ${\cal Z}^{(+-)}_{++}$ implies,
roughly speaking, that the two
$\widehat{gL}(n)$'s should be identified. This reduces the $N=2$
superfield degrees of freedom carried by $Q^{\pm (0)}_+$ from $2(n-1)$
down to $(n-1)$. This is the desired number required by the
$W\hat A(n-1\vert n-1)^{(1)}$ Miura transformation.

Now writing the ${\cal Z}^{(+-)}_{++}$ conformal operator as
$${\cal Z}^{[+-]}_{++} =\tau_{++}\ominus\tau_{++}\ ,           \eqno(4.47)$$
where $\tau_{++}$ is a conformal spin one operator valued in the
$\widehat{gL}(n)^{(1)}/gL(1)$ Lie algebra. Then using Eqs.(4.42)--(4.45)
and Eqs.(2.35), (2.36), we get
$$\tau_{++}=\nabla_{++}+J_{++}+\lambda_{++}                       \eqno(4.48)$$
together with the $gL(1)$ factorization condition
$$<n\vert J_{++}\vert n>\ =\ <2n\vert J_{++}\vert 2n>\ =0\ .$$

This object has the same structure as the $N=0$ Lax operator Eqs.(4.1)
and (4.3). The role of $\partial_{++}$ and $q_{++}$ are played here by
$\nabla_{++}$ and $J_{++}$ respectively. The only difference is that
$tr(J_{++})$ is no longer zero.

Furthermore, applying the analysis of Subsection 4.1 to the $\tau_{++}$
operator, one gets the manifestly $N=2$ superfield Miura transformation.
It reads as
$$\left(\nabla^n_{++}+\sum^{n-1}_1\ T_k\ \nabla^{n-k}_{++}\right) =
\left[\mathop{\pi}\limits^{n-1}_{j=1}\ (\nabla_{++}-J_{++})_j\right]
\nabla_{++}
        \eqno(4.49)$$
where
$$(\nabla_{++}-J_{++})_j=\nabla_{++}-J_{++j}\ .
\eqno(4.50)$$
Here $J_{++j}$ denotes the $j$--th component of the $J_{++}$ $n\times n$
matrix calculated in the diagonal gauge $Q^{\pm (0)}_+$. It is given by
$$\eqalignno{
  J_{++j} &=\ <j\vert J^{(0)}_{++}\vert j>\cr
  &= D^+_+Q^-_{+j}-D^-_+Q^+_{+j}+2Q^+_{+j}Q^-_{Ij}\ .       &(4.51)\cr}$$
Expanding the r.h.s. of Eq.(4.49), one gets the $N=2$ superfield of the
$W_s$ supercurrents; $1\leq s\leq n-1$. For the $n=2$ case, we have
one conserved conformal spin one supercurrent $T_1$. It is equal to
$J_{++}$. For $n=3$, we have two conserved supercurrents $T_1$ and
$T_2$ having spin one and two respectively. These $N=2$ supercurrents,
which are built out of two $N=2$ chiral superfields $\psi^\pm_+$ and
$\highchi^\pm_+$ or equivalently $(J_{++})_1$ and $(J_{++})_2$, read
as
$$\eqalign{
  T_1 &= -(J_{++})_1-(J_{++})_2\cr
  T_2 &= (J_{++})_i\cdot (J_{++})_2-\nabla_{++} (J_{++})_2\ .\cr} \eqno(4.52)$$

In the next section, we shall develop another method, based on the
analysis of subsection 4.2 and the $U(1)$ charge decomposition, to
construct $N=2$ superfield realizations of the $N=2$ $W_s$ supercurrents.

\SECTION{$N=2$\ $W_s$ CURRENTS: THE $U(1)$ CHARGE
DECOMPOSTION METHOD}

Field realization of higher conformal spin currents $W_s$ are easily
derived once we know the Miura transformation. The manifestly
$N=2$ formula we derived so far is based on the existence of a solution
of the constraint Eqs.(4.49). Here we develop an alternative method
leading to $N=2$ superfield realization of the $W_s$ currents by using
Eq.(4.32). The idea is to start from a manifestly $N=2$ formalism. Then
break $N=2$ supersymmetry down to $N=1$. Finally, truncate the $N=1$
results by keeping the zero $U(1)$ charge contributions. This gives the
desired $N=2$ quantities.

To start, consider the $N=2\ U(1)$ Lax operators of Eqs.(4.33) and
define the new ones
$$\eqalignno{
  {\cal D}_+ &={\cal D}^+_++{\cal D}^-_+ &(5.1)\cr
  \bar{\cal D}_+ &={\cal D}^+_+-{\cal D}^-_+  &(5.2)\cr}$$
obeying the superalgebra
$$\eqalignno{
  \{{\cal D}_+,{\cal D}_+\} &=\{{\cal D}^+_+,{\cal D}^-_+\} &(5.3)\cr
  \{\bar{\cal D}_+,\bar{\cal D}_+\} &=-\{{\cal D}^+_+,{\cal D}^-_+\} &(5.4)\cr
  \{\bar{\cal D}_+,{\cal D}_+\} &=[{\cal D}^+_+,{\cal D}^-_+]\ .   &(5.5)\cr}$$
Then, we focus our attention on ${\cal D}_+$ only. Naturally this procedure
breaks $N=2$ supersymmetry down to $N=1$. Using Eqs.(4.33),
${\cal D}_+$ may be rewritten as
$${\cal D}_+=D_++Q_++\Lambda_+
\eqno(5.6)$$
where
$$\eqalign{
  D_+ &=D^+_++D^-_+\cr
  Q_+ &=Q^+_++Q^-_+\cr
  \Lambda_+ &=\Lambda^+_++\Lambda^-_+\cr}                          \eqno(5.7)$$
Eqs.(5.6) and (5.7) give a special representation of the $N=1$
supersymmetric Lax operator Eq.(4.16).

Applying the analysis of Subsection 4.2 to this operator, one gets the
$W\hat A(n-1\vert n-1)^{(1)}$ Miura transformation, see Eqs.(4.28)--(4.32)
$$D^{2n}_++\sum^{2n}_{i=2}\ U_i\ D^{2n-i}_+=
  \mathop{\Pi}\limits^n_{j=1}\ (D_+-\xi_+)_j
  (D_+-\xi_+)_{j+n}\ ,
    \eqno(5.8)$$
where $D_+$ is given by Eq.(5.7). Moreover, since the $\xi_+$'s carry
$2(n-1)$ real degree of freedom as required by Eq.(4.23), we use
complex notations by setting
$$\eqalign{
  (\xi_+)_{2j-1} &=\alpha_j(\eta^+_+)_j+\bar\alpha_j
  (\eta^-_+)_j\cr
  (\xi_+)_{2j} &=-i\left[\beta_j(\eta^+_+)_j-\bar\beta_j
  (\eta^-_+)_j\right]\cr}
     \eqno(5.9)$$
where $\alpha_j, j=1,\dots n-1$, are $\bbc$ numbers and where
$(\eta^+_+)_j$ and $(\eta^-_+)_j$ are $N=2$ chiral and antichiral
superfields obeying:
$$D^+_+(\eta^+_+)_j=D^-_+(\eta^-_+)_j=0\ .
\eqno(5.10)$$
Plugging Eqs.(5.9) back into Eq.(5.8), one gets the $W\hat A(n-1\vert
n-1)^{(1)}$
Miura transformation in terms of the $N=2$ superfields. Expanding the
r.h.s., one finds that the higher conformal spin currents $U_k$ have the
following $U(1)$ charge decomposition:
$$\eqalign{
  U_{2k} &=\kappa^k\ J^0_k+(J^{++}_k+J^{--}_k)\cr
  U_{2k+1} &=\kappa^k\left(\Gamma^+_{k+{1\over 2}}+
  \Gamma^-_{k+{1\over 2}}\right)\cr}                              \eqno(5.11)$$
where $\kappa$ is a constant which may be determined by normalization
conditions of the $N=2\ U(1)$ supercurrent. Higher $U(1)$ representations
in Eq.(5.11) are forbidden by the $N=2$ chirality Eqs.(5.10) and the
Grassmann structure of fermions. Moreover, the reality condition of the
$U_k$'s implies
$$\eqalign{
  (J^0_k)^* &=J^0_k\cr
  (J^{--}_k)^* &=J^{++}_k,\ \ \ \Gamma^-_{k+{1\over 2}}=
  \left(\Gamma^+_{k+{1\over 2}}\right)^*\ .\cr}
\eqno(5.12)$$
The $U(1)$ charge decomposition method classifies the $U_k$
supercurrents in five blocks $J^0_k, \Gamma^+_{k+{1\over 2}},
\Gamma^-_{k+{1\over 2}}$,
$J^{++}_k$ and $J^{--}_k$. $J^0_k$ is the only hermitian block. Knowing
that $1\leq k\leq n-1$, one sees that for a given $n$ there are exactly
$(n-1)$ conserved $N=2$ hermitian supercurrents. This property
together with $U(1)$ covariance ensure that $J^0_k$ are indeed the
$N=2$ $W$ supercurrents described earlier by Eqs.(4.23)--(4.25). In
what follows we illustrate this idea with examples.

\SECTION{EXAMPLE ONE}

Here we give the full structure of the $U(1)$ charge decomposition
of the $W\hat A(1\vert 1)^{(1)}$ Miura transformation. As a consequence
we obtain the $N=2$ superfield realizations of the usual hermitian
$N=2$ current $J^0_{++}\equiv J^0_1$ and a complex conformal spin
${3\over 2}$ supercurrent $F^+_{+++}=F^+_{3/2}$. The component field
content of $J^0_1$ and $F^+_{3/2}$ are, respectively, given by
$(j^0_1(z),g^\pm_{3/2}(z),t^0_2(z))$ and $(f^+_{3/2}(z), h^+_2(z), h^{++}_2(z),
k^+_{5/2}(z))$. The lower index carried by the fields indicate their
conformal spin weight.

We begin by recalling that the diagonal and canonical gauges of the
$Q_+$ superfield read respectively as:
$$\eqalignno{
  Q^{(0)}_+ &=\sum^2_{j=1}\left(\xi_{+j}\vert j><j\vert +
  \xi_{+j+n}\vert j+n><j+n\vert\right) &(6.1)\cr
  Q^{(1)}_+ &=\sum^2_{j=1}\left( U_{2j-1}\vert 2j-1><1\vert +
  U_{2j}\vert 2j><1\vert\right)\ ,   &(6.2)\cr}$$
where the $U_k$ superfields behave as ${k\over 2}$ conformal spin
objects. As the supertrace is vector basis independent, we have
$$Str\ Q^{(0)}_+ = Str\ Q^{(1)}_+\ .
\eqno(6.3)$$
The $SL(2\vert 2)$ vanishing supertrace condition implies
$$\eqalignno{
  \xi_{+4}+\xi_{+3} &=\xi_{+1}+\xi_{+2} &(6.4)\cr
  U_1 &= 0\ , &(6.5)\cr}$$
leaving then three real superfield variables only. Making the choice
$\xi_{+4}=0$, which in turn leads to take $U_4=0$, the number of
degrees of freedom reduces to two as shown by the following equation
$$\eqalign{
  \xi_{+4} &=0,\xi_{+3}=\xi_{+1}+\xi_{+2}\cr
  U_1 &=U_4=0\ .\cr}
\eqno(6.6)$$
The $\hat A(1\vert 1)^{(1)}$ Miura transformation then reads as
$$D^4_++U_2D^2_++U_3D^3_+=(D_+-\xi_{+1})(D_+-\xi_{+3})(D_+-
  \xi_{+2})D_+\ .
        \eqno(6.7)$$
Expanding the r.h.s. of Eq.(6.7) and using Eqs.(6.6) we find after some easy
algebra
$$\eqalignno{
  U_2 &= -D_+(\xi_{+1}+\xi_{+2})+\xi_{+1}\xi_{+2} &(6.8)\cr
  U_3 &=-2\partial_{++}\xi_{+2}+\xi_{+2}D_+\xi_{+1}\ . &(6.9)\cr}$$
To make contact with the $N=2$ language we take $\xi_{+i}$ as
$$\eqalign{
  \xi_{+1} &=\alpha_1\psi^+_++\bar\alpha_1\psi^-_+\cr
  \xi_{+2} &=-i(\alpha_2\psi^+_+-\bar\alpha_2\psi^-_+)\ ,\cr}  \eqno(6.10)$$
where $\alpha_1$ and $\alpha_2$ are two complex parameters and
where $\psi^+_+$ and $\psi^-_+=(\psi^+_+)^*$ are $N=2$ chiral and
antichiral superfields respectively. They read in terms of $\xi_{+1}$
and $\xi_{+2}$ as
$$\eqalign{
  \psi^+_+ &= {1\over\alpha_1\alpha_2}\ [\alpha_2\xi_{+1}+
  i\alpha_1\xi_{+2}]\cr
  \psi^-_+ &={1\over\bar\alpha_1\bar\alpha_2}\ [\bar\alpha_2
  \xi_{+1}-i\bar\alpha_1\xi_{+2}]\ .\cr}
\eqno(6.11)$$
Putting back Eqs.(6.10) into (6.8), (6.9), we get the $U(1)$ charge
decomposition of the $N=2$ conformal spin 1 and ${3\over 2}$ currents.

\SUBSECTION{The $U_2$ supercurrents}

First of all note that from Eqs.(6.8) and (6.10), one learns that the
general $U(1)$ charge decomposition of the conformal spin 1 supercurrent
reads as
$$U_2=i\kappa\ J^{[+-]}_{++}+\tilde\kappa\ J^{(+-)}_{++}+
  y\ J^{++}_{++}+\bar y\ J^{--}_{++}\ ,        \eqno(6.12)$$
where $\kappa ,\tilde\kappa ,y$ and $\bar y$ are $\bbc$
numbers to be determined
later on. The $U_2$ reality condition implies
$$\eqalign{
  \left( J^{[+-]}_{++}\right)^* &=-J^{[+-]}_{++}\cr
  J^{(+-)}_{++} &= J^{(+-)}_{++}\cr
  J^{--}_{++} &= (J^{++}_{++})^*\cr
  \kappa^* &=\kappa ,\tilde\kappa^*=\tilde\kappa\ .\cr}      \eqno(6.13)$$

To find the $N=2$ superfield realization of the irreducible terms of
Eq.(6.12), we proceed as follows. First, calculate $(\xi_{+1}+\xi_{+2})$
and $\xi_{+1}\cdot\xi_{+2}$ by using Eqs.(6.10). We get
$$\eqalign{
  \xi_{+1}+\xi_{+2} &=-i(\alpha_2+i\alpha_1)\psi^+_++i(\bar\alpha_2-
  i\bar\alpha_1)\psi^-_+\cr
  \xi_{+1}\cdot\xi_{+2} &=i(\alpha_1\bar\alpha_2+\bar\alpha_1\alpha_2)
  \psi^+_+\psi^-_+\ .\cr}
     \eqno(6.14)$$
Putting back into Eq.(6.8), the supercurrent $U_2$ takes the form
$$U_2=-i\sqrt 2\ (\alpha_1\bar\alpha_2+\bar\alpha_1\alpha_2)
  \left[ +{1\over\sqrt 2}\ \psi^+_+\psi^-_++\alpha_0D^-_+\psi^+_+-
  \bar\alpha_0D^+_+\psi^-_+\right]\ ,
\eqno(6.15)$$
where
$$\alpha_0={1\over\sqrt 2}\ {(\alpha_2+i\alpha_1)\over
  \alpha_1\bar\alpha_2+\bar\alpha_1\alpha_2}\ .                   \eqno(6.16)$$
Comparing with the decomposition Eq.(6.12), one finds that the $\tilde\kappa$
and $y$ $\bbc$ numbers are identically zero and $\kappa =\sqrt 2(\alpha_1
\bar\alpha_2+\bar\alpha_1\alpha_2)$. Thus, the nonvanishing term
namely
$$J^{[+-]}_{++} = +{1\over\sqrt 2}\ \psi^+_+\psi^-_++\bar\alpha_0
  D^+_+\psi^-_+-\alpha_0D^-_+\psi^+_-
\eqno(6.17)$$
is just the general $N=2$ superfield realization obtained in Ref.[12].
Setting
$$\eqalign{
  \psi^-_+ &= D^-_+\phi,\quad D^-_+\phi =0\cr
  \psi^+_+ &=D^+_+\bar\phi ,\quad D^+_+\bar\phi =0\ ,\cr}    \eqno(6.18)$$
then plugging back into Eq.(6.17), one recovers the Zheng and Yu
realization [25, 12].

Finally, note that using the following two points functions,
$$\eqalign{
  <\psi^+_+(Z_1)\psi^-_+(Z_2)>\ &=-{1\over Z_{12}}+
  {\theta^+_{+12}\theta^-_{+12}\over Z^2_{12}}\cr
  <\psi^+_+(Z_1)\psi^+_+(Z_2) &=\ <\psi^-_+(Z_1)\psi^-_+(Z_2)>\ =0\cr}

             \eqno(6.19)$$
with $\theta^\pm_{+12}=\theta^\pm_{+1}-\theta^\pm_{+2}$ and
$Z_{12}=z_{12}+(\theta^+_{+1}\theta^-_{+2}+\theta^-_{+1}\theta^+_{+2})$,
one may check that the quantum version of Eq.(6.17) satisfy the O.P.E.
algebra
$$\eqalignno{
  J^{[+-]}_{++}(Z_1)J^{[+-]}_{++}(Z_2) &={c_2\over 2Z^2_{12}}+
  {1\over 2Z_{12}}\left[\theta^+_{+12}D^-_{+2}-\theta^-_{+12}
  D^+_{+2}+{\theta^+_{+12}\theta^-_{+12}\over Z_{12}}\
  \partial_{++12}\right] J^{(+-)}_{++}(Z_2)\cr
  &\quad +{\theta^+_{+12}\theta^-_{+12}\over Z^2_{12}}\
  J^{[+-]}_{++}(Z_2)\ .
         &(6.20)\cr}$$
where the central charge $c_2$ is equal to $(1+8\vert\alpha_0\vert^2)$.

\SUBSECTION{The $U_3$ supercurrent}

This is a real conformal spin ${3\over 2}$ supercurrent which in the
$N=1$ CFT's describes just the $N=1$ super energy momentum tensor.
In the $N=2$ formalism, this object turns out to be the real part of a
complex conformal spin ${3\over 2}\ N=2$ current. Indeed, from Eqs.(6.9),
(6.10) and the $N=2$ chirality property of $\psi^+_+$ Eq.(5.10), one
learns that $U_3$ has the following $U(1)$ charge decomposition:
$$U_3=i\ \kappa [F^*_{+++}-F^-_{+++}]\ ,
\eqno(6.21)$$
where $F^-_{+++}$ is the complex conjugate of $F^+_{+++}$ and $\kappa$
is the same as in Eq.(6.12). Using Eqs.(6.10) and Eqs.(5.7)--(5.10), we
calculate
$$\eqalign{
  \xi_{+2}D_+\xi_{+1} &=-i\bar\alpha_1\alpha_2\psi^+D^+_+\psi^-_+-
  i\alpha_1\bar\alpha_2\psi^+_+D^-_+\psi^+_++c.c\cr
  -\partial_{++}\xi_{+2} &= i\alpha_2\partial_{++}\psi^+_++c.c\cr}\eqno(6.22)$$
Plugging back into Eq.(6.9), we find the $N=2$ superfield realization of
$F^+_{+++}$. It reads as
$$F^+_{+++}=-{\alpha_1\bar\alpha_2\over\kappa}\ \psi^+_+D^-_+
  \psi^+_+-{\bar\alpha_1\alpha_2\over\kappa}\ \psi^+_+ D^+_+
  \psi^-_++{2\alpha_2\over\kappa}\ \partial_{++}\psi^+_+\ .   \eqno(6.23)$$
Expanding this object in $\theta^\pm_+$ series as
$$F^+_{3/2}=f^+_{3/2}+\theta^+_-h^{-+}_2+\theta^-_-h^{++}_2+
  \theta^+_-\theta^-_-k^+_{5/2}\ ,
\eqno(6.24)$$
one sees the $F^+_{3/2}$ multiplet consists of: Two $U(1)$ complex
vectors of spin ${3\over 2}$ and ${5\over 2}$, a conformal spin
$2\ U(1)$ complex scalar and a $U(1)$ rank two tensor of spin 2.
Putting $J^{[+-]}_{++}$ and $F^\pm_{++}$ altogether one expects that the
$N=2$ superfield realizations given by Eqs.(6.17) and (6.23) generate
a $N=2$ higher conformal spin algebra involving, besides the usual
$N=2\ U(1)$ conformal currents, namely $\left( j_1, g^\pm_{3/2},t_2
\right)$, the analytic currents $\left( f^\pm_{3/2},h^{+-}_2,\bar h^{+-}_2,
h^{++}_2,h^{--}_2,k^\pm_{5/2}\right)$. As the conformal spin of
$F^+_{3/2}$ is $3/2$, one may refer to this algebra as the $N=2\ W_{3/2}$
algebra. This is a complex $N=2\ W$ algebra and has no physical
application. It would be interesting to write down the O.P.E. of this
nonlinear algebra.

\SECTION{EXAMPLE TWO}

This example deals with the $U(1)$ charge decomposition of the
$\hat A(2\vert 2)^{(1)}$ Miura transformation. Solving the $\hat A(2\vert
2)^{(1)}$ consistency condition Eq.(4.23) in terms of two $N=2$ chiral
superfields $\psi^+_+$ and $\highchi^+_+$ and their complex conjugates,
we obtain a $N=2$ superfield realizations of the $N=2$ conformal spin 1,
${3\over 2}$ and 2 supercurrents respectively. Moreover, we find that
the $N=2\ U(1)$ conformal spin 1 current $J^{[+-]}_{++}$ has the form
$$J_{++}=J^\psi_{++}+J^\chi_{++}+J^{\psi\chi}_{++}\ ,
\eqno(7.1)$$
where $J^\psi_{++}$ and $J^\chi_{++}$ are the usual $N=2$ currents
associated with the superfields $\psi^\pm_+$ and $\highchi^\pm_+$
respectively and where the coupling term has exactly the property of
a Feigin Fuchs extension.

We start by giving the $\widehat{SL}(3\vert 3)^{(1)}$ diagonal and
canonical form of the spinor superfield $Q_+$
$$\eqalignno{
  Q^{(0)}_+ &=\sum^3_{j=1}\ \left(\xi_{+j}\vert j><j\vert +
  (\xi_+)_{j+3}\vert j+3><j+3\vert\right) &(7.2)\cr
  Q^{(1)}_+ &=\sum^3_{j=1}\left(U_{2j-1}\vert 2j-1><1\vert +
  U_{2j}\vert 2j><1\vert\right)\ .   &(7.3)\cr}$$
The $SL(3\vert 3)$ vanishing supertrace condition implies
$$\eqalignno{
  \xi_{+6}+\xi_{+5}+\xi_{+4} &=\xi_{+1}+\xi_{+2}+\xi_{+3}\cr
  U_1 &= 0\ .                              &(7.4)\cr}$$

The extra condition $\xi_{+6}=0$ and equivalently $U_6=0$, allows us
to solve Eq.(8.4) by four real superfield variables as in Eqs.(4.30) and
(4.31) namely:
$$\eqalign{
  \xi_{+1} &=\psi_{+1}\cr
  \xi_{+2} &=\psi_{+2}+\psi_{+3}\cr
  \xi_{+3} &=\psi_{+4}\cr
  \xi_{+4} &=\psi_{+1}+\psi_{+2}\cr
  \xi_{+5} &=\psi_{+3}+\psi_{+4}\ .\cr}
\eqno(7.5)$$
The $W\hat A(2\vert 2)^{(1)}$ Miura transformation reads as
$$D^6_++\sum^5_{j=2}\ U_jD^{6-j}_+=(D_+-\xi_{+1}(D_+-\xi_{+4})
  (D_+-\xi_{+2})(D_+-\xi_{+5})(D_+-\xi_{+3})D\ .                  \eqno(7.6)$$

Expanding the r.h.s. of this operator equation and using Eqs.(7.5), one
finds after a lengthy but straightforward algebra, see Appendix A.
$$\eqalignno{
  U_2 &=\left[ -D_+(\psi_{+1}+\psi_{+2})+\psi_{+1}\psi_{+2}\right] +
  \left[ -D_+(\psi_{+3}+\psi_{+4})+\psi_{+3}\psi_{+4}\right] +
  \psi_{+2}\psi_{+3} &(7.7)\cr
  &\cr
  U_3 &= -D^2_+(\psi_{+2}+\psi_{+4})+D_+(\psi_{+1}\psi_{+2})+
  D_+(\psi_{+3}\psi_{+4})+\psi_{+1}D_+\psi_{+2}-\psi_{+2}
  D_+\psi_{+3}\cr
  &\quad -\psi_{+2}D_+\psi_{+4}+\psi_{+2}\psi_{+3}\psi_{+4} &(7.8)\cr
  &\cr
  U_4 &= -D^3_+(\psi_{+3}+\psi_{+4})+D^2_+(\psi_{+2}\psi_{+3} +
  \psi_{+3}\psi_{+4})+D_+\left( (\psi_{+1}+\psi_{+2})D_+
  (\psi_{+3}+\psi_{+4})\right)\cr
  &\quad -(\psi_{+1}+\psi_{+2})D_+(\psi_{+3}\psi_{+4})+
  (\psi_{+1}+\psi_{+2})D^2_+\psi_{+4}\cr
  &\quad -D_+ (\psi_{+1}\psi_{+2}\psi_{+3} +
  \psi_{+1}\psi_{+3}\psi_{+4}+\psi_{+2}\psi_{+3}\psi_{+4}) &(7.9)\cr
  &\quad +\psi_{+1}D^2_+\psi_{+3}-\psi_{+1}D_+(\psi_{+2}\psi_{+3})-
  \psi_{+1}\psi_{+2}D_+(\psi_{+3}+\psi_{+4})+\psi_{+1}\psi_{+2}
  \psi_{+3}\psi_{+4}\cr
\noalign{\hbox{and}}
  U_5 &= -D^4_+\psi_{+4}+D^3_+(\psi_{+3}\psi_{+4})-D_+\left[
  (\psi_{+1}+\psi_{+2})\left( D_+ (\psi_{+3}+\psi_{+4}) -
  D^2_+\psi_{+4}\right)\right]\cr
  &\quad +\psi_{+1}D^3_+\psi_{+3}+\psi_{+1}D^2_+(\psi_{+1}\psi_{+3})
  +\psi_{+1}\psi_{+2}
  \left[ D_+ (\psi_{+3}+\psi_{+4}) -D^2_+\psi_{+4}\right]\ .   &(7.10)\cr}$$

\SUBSECTION{The $U_2$ supercurrent}

The $U(1)$ charge decomposition of the $N=2$ conformal spin 1
current follows by making the following change of variables generalizing
the $SL(2\vert 2)$ one Eqs.(6.10)
$$\eqalign{
  \psi_{+1} &=\alpha_1\psi^+_++\bar\alpha_1\psi^-_+\cr
  \psi_{+2} &= -i(\alpha_2\psi^+_+-\bar\alpha_2\psi^-_+)\cr
  \psi_{+3} &=\alpha_3\highchi^+_++\bar\alpha_3\highchi^-_+\cr
  \psi_{+4} &= -i(\alpha_4\highchi^+_+-\bar\alpha_4
  \highchi^-_+)\ ,\cr}
    \eqno(7.11)$$
or equivalently
$$\eqalign{
  \psi^+_+ &= {1\over \alpha_1\alpha_2}\ (\alpha_2\psi_{+1} +
  i\alpha_1\psi_{+2})\cr
  \highchi^+_+ &+ {1\over \alpha_3\alpha_4}\ (\alpha_4\psi_{+3} +
  i\alpha_3\psi_{+4})\cr
  \psi^-_+ &= (\psi^+_+)^*;\ \  \highchi^-_+=(\highchi^+_+)^*\
.\cr}\eqno(7.12)$$
The $\alpha_i$'s are $\bbc$ numbers. They play the same role as in
the $\hat A(1\vert 1)^{(1)}$ case, see Eqs.(6.17) and (6.20).

Using Eqs.(7.11), we calculate
$$\eqalign{
  \psi_{+1}+\psi_{+2} &= (\alpha_1-i\alpha_2)\psi^+_+ +(\bar\alpha_1+
  i\bar\alpha_2)\psi^-_+\cr
  \psi_{+3}+\psi_{+4} &= (\alpha_3-i\alpha_4)\highchi^+_+ +
  (\bar\alpha_3+i\bar\alpha_4)\highchi^-_+\cr
  \psi_{+1}\psi_{+2} &= i(\alpha_1\bar\alpha_2+\bar\alpha_1
  \alpha_2)\psi^+_+\psi^-_+\cr
  \psi_{+3}\psi_{+4} &= i(\alpha_3\bar\alpha_4+\bar\alpha_3
  \alpha_4)\highchi^+_+\highchi^-_+\cr
  \psi_{+2}\psi_{+3} &= i(\alpha_2\bar\alpha_3\psi^+_+
  \highchi^-_+ +\bar\alpha_2\alpha_3\highchi^+_+
  \psi^-_+)-i\alpha_2\alpha_3\psi^+_+\highchi^+_++
  i\bar\alpha_2\bar\alpha_3\psi^-_+\highchi^-_+\ .\cr}      \eqno(7.13)$$

Putting back into Eq.(7.7) and taking into account (5.7), the $U_2$
supercurrent reads as
$$U_2=U^\psi_2+U^\chi_2+U^{\psi\chi}_2\ ,
\eqno(7.14)$$
where $U^\psi_2$ is given by Eqs.(6.15)--(6.17) and where
$$U^\chi_2=i\sqrt 2\ (\alpha_3\bar\alpha_4+\bar\alpha_3
  \alpha_4)\left[{1\over\sqrt 2}\ \highchi^+_+\highchi^-_+ +
  \beta_0D^-_+\highchi^+_+-\bar\beta_0D^+_+\highchi^-_+\right] \eqno(7.15)$$
with
$$\beta_0={(\alpha_4+i\alpha_3)\over\sqrt 2\ (\alpha_3
  \bar\alpha_4+\bar\alpha_3\alpha_4)}\ .                          \eqno(7.16)$$
The last term $U^{\psi\chi}_2$, which describes the interactions between
the $\psi^\pm_+$ and $\highchi^\pm_+\ N=2$ superfields, has the
following remarkable $U(1)$ charge decomposition
$$U^{\psi\chi}_2 =i\kappa\left( J^{-+}_{++} +J^{++}_{++} +
  J^{--}_{++}\right)\ ,
   \eqno(7.17)$$
where
$$J^{-+}_{++} =\highgamma_0\highchi^-_+\psi^+_+-\bar{\highgamma}_0
  \highchi^+_+\psi^-_+
    \eqno(7.18)$$
and
$$\eqalign{
  J^{++}_{++} &=\delta_0\highchi^+_+\psi^+_+;\ \  J^{--}_{++} =
  \bar\delta_0\highchi^-_+\psi^-_+\cr
  (J^{++}_{++})^2 &= 0,\quad (J^{--}_{++})^2 = 0\ .\cr}
                                                            \eqno(7.19)$$
$\highgamma_0$ and $\delta_0$ are related to the $\alpha_i$'s as
$$\eqalign{
  \highgamma_0 &={\alpha_2\bar\alpha_3\over\kappa},\quad
  \delta_0 ={\alpha_2\alpha_3\over\kappa}\cr
  \kappa &= \sqrt 2\ (\alpha_1\bar\alpha_2+\bar\alpha_1
  \alpha_2)=\sqrt 2\ (\alpha_3\bar\alpha_4 +\bar\alpha_3\alpha_4)\ .\cr}

               \eqno(7.20)$$
The second relation of Eqs.(7.20) is imposed in order to have the same
normalization for the $\psi^\pm_+$ and $\highchi^\pm_+$ superfields.
In particular we have
$$\eqalign{
  <\highchi^+_+(Z_1)\highchi^-_+(Z_2)>\ &=-Z^{-1}_{12}+Z^{-2}_{12}
  \theta^+_{+12}\theta^-_{+12}\cr
  <\highchi^+_+(Z_1)\highchi^+_+(Z_2)>\ &=\ <\highchi^-_+(Z_1)
  \highchi^-_+(Z_2).\ =0\ .\cr}
  \eqno(7.21)$$
Using the $U(1)$ charge decomposition of $U_2$ Eq.(7.7) and taking
into account Eqs.(7.12)--(7.14) and (7.15)--(7.18), we get the following
$N=2$ superfield realization of the $N=2$ current
$$\eqalign{
  J^{[+-]}_{++} &= \left[{1\over\sqrt 2}\ \psi^+_+\psi^-_+ +
  \alpha_0D^-_+\psi^+_+-\bar\alpha_0D^+_+\psi^-_+\right]\cr
  &\quad +\left[{1\over\sqrt 2}\ \highchi^+_+\highchi^-_+ +
  \beta_0D^-_+\highchi^+_+-\bar\beta_0D^+_+\highchi^-_+\right]\cr
  &\quad +\left[ 0+\highgamma_0\highchi^-_+\psi^+_+-\bar{\highgamma}_0
  \highchi^+_+\psi^-_+\right]\ .\cr}
 \eqno(7.22)$$
Eq.(7.22) contains, in addition to the $\psi^\pm$ and $\highchi^\pm$
supercurrents, an extra piece which may be viewed as a Feigin Fuchs
extension of the $J(\psi )$ and $J(\highchi)$ current. Using Eqs.(6.19) and
(7.21), one sees that these terms contribute to the central charge $c_2$
in the same way as do the terms $\alpha_0D^-_+\psi^+_+$ and
$\beta_0D^-_+\highchi^+_+$. This can be easily seen by help of the
identity
$$<\psi^+_+(1)\psi^-_+(2)><\highchi^-_+(1)\highchi^+_+(2)>\ =
  Z^{-2}_k\left( 1-Z^{-1}_{12}\theta^+_{12}\theta^-_{12}\right)
  \left( 1+Z^{-1}_{12}\theta^+_{12}\theta^-_{12}\right) =Z^{-2}_{12}\ .

               \eqno(7.23)$$

Note finally, that the realization (7.22) depends on the $(3+3)$ $\bbc$
numbers $\alpha_0,\beta_0,\highgamma_0$ and their complex conjugates.
Setting $B_0=\highgamma_0=0$, one recovers the realization of Ref.[16].

\SUBSECTION{The $U_3$ supercurrent}

Following the same analysis as done before, one finds that the
$U_3$ object of Eq.(7.8) is in fact the real part of a $N=2$ complex
conformal spin ${3\over 2}$ current $K^+_{3/2}$. Using Eqs.(7.11) and (5.7),
the $U_3$ hermitian superfield reads as
$$U_3=i\kappa (K^+_{+++}-K^-_{+++})\ ,
\eqno(7.24)$$
where $K^-_{+++}=(K^+_{+++})^*$ and where $K^+_{+++}$ has the following
$N=2$ superfield realization, see Appendix B.
$$\eqalignno{
  K^+_{+++} &= D^+_+\left[\psi^+_+\psi^-_++\highchi^+_+\highchi^-_++
  {\alpha_2\over\kappa}\ D^-_+\psi^+_++{\alpha_4\over\kappa}\
  D^-_+\highchi^+_+\right]\cr
  &\quad +\left[{\alpha_1\bar\alpha_2\over\kappa}\ \psi^+_+
  D^+_+\psi^-_+-{\alpha_1\alpha_2\over\kappa} \psi^+_+D^-_+
  \psi^+_++{\alpha_2(\bar\alpha_3+i\bar\alpha_4)\over\kappa}\
  \psi^+_+D^+_+\highchi^-_+\right]\cr
  &\quad +\left[{\alpha_2\alpha_3\over\kappa}\ \psi^+_+D^-_+\highchi^+_+
  -{i\alpha_2\over\kappa}\ \psi^+_+\highchi^+_+\highchi^-_+ -
  {\alpha_2(\alpha_3-i\alpha_4)\over\kappa}\ \psi^+_+ D^-_+
  \highchi^+_+\right]\ .
       &(7.25)\cr}$$

\SUBSECTION{The $U_4$ supercurrent}

Similar analysis shows that the $U(1)$ charge decomposition of $U_4$
reads as
$$U_4=-\kappa^2T_{++++}+\kappa L^{++}_{++++}+\kappa L^{--}_{++++}\ ,

              \eqno(7.26)$$
where
$$\eqalign{
  T^*_{++++} &= T_{++++}\cr
  L^{--}_{++++} &=(L^{++}_{++++})^*\ .\cr}
\eqno(7.27)$$
To work out the $N=2$ superfield realization of the conformal spin 2
supercurrent $T_{++++}$, one uses Eqs.(5.7) and (7.11)--(7.26). As the
$N=2$ super $W_3$ currents have been considered recently, we shall
give hereafter the main steps of our calculations in order to compare
the obtained results. We have:
$$-D^3_+(\psi_{+3}+\psi_{+4})=-2(\alpha_3-i\alpha_4)\partial_{++}
  D^-_+\highchi^+_+ +c.c
\eqno(7.28)$$
$$-D^2_+(\psi_{+3}+\psi_{+4}) = -2(\alpha_3-i\alpha_4)
  \partial_{++} D^-_+\highchi^+_+ +c.c
\eqno(7.29)$$
$$D^2_+(\psi_{+3}\psi_{+4}+\psi_{+2}\psi_{+3})=2i\kappa\
  \partial_{++} (\highchi^+_+\highchi^-_+) -2i\alpha_2\bar\alpha_3
  \partial_{++}(\psi^+_+\highchi^-_+) -2i\alpha_2\alpha_3
  \partial_{++} (\psi^+_+\highchi^+_+) +c.c
\eqno(7.30)$$
$$\eqalignno{
  D_+\left[ (\psi_{+1}+\psi_{+2})D_+(\psi_{+3}+\psi_{+4}\right] &=
  (\alpha_1-i\alpha_2)(\alpha_3-i\alpha_4)\left[ D^-_+\psi^+_+
  D^-_+\highchi^+_+ -2\psi^+_+\partial_{++}\highchi^-_+\right] +c.c\cr
  &\quad +(\alpha_1-i\alpha_2)(\bar\alpha_3+i\bar\alpha_4)
  \left[ D^-_+\psi^+_+D^+_+\highchi^-_+-2\psi^+_+\partial_{++}
  \highchi^-_+\right] + c.c\cr
  & &(7.31)\cr}$$
$$-(\psi_{+1}+\psi_{+2})D_+(\psi_{+3}\psi_{+4})=-i\kappa (\alpha_1-
  i\alpha_2)\left[\psi^+_+\highchi^-_+D^-_+\highchi^+_+-
  \psi^+_+\highchi^+_+D^+_+\highchi^-_+\right] +c.c
\eqno(7.32)$$
$$\eqalignno{
  (\psi_{+1}+\psi_{+2})D^2_+\psi_{+4} &= -2i\alpha_4(\alpha_1-i\alpha_2)
  \psi^+_+\partial_{++}\highchi^+_+ +c.c\cr
  &\quad -2i\alpha_4(\bar\alpha_1+i\bar\alpha_2)\psi^-_+
  \partial_{++}\highchi^-_+ +c.c
     &(7.33)\cr}$$
$$\eqalignno{
  -D_+(\psi_{+1}\psi_{+2}\psi_{+3}+\psi_{+1}\psi_{+3}\psi_{+4}+
  \psi_{+2}\psi_{+3}\psi_{+4}) &= i\kappa\left[\alpha_3\psi^-_+
  D^-_+(\psi^+_+\highchi^+_+)\right] + c.c\cr
  &\quad +i\kappa\left[ (\alpha_1-i\alpha_2)\highchi^-_+D^-_+
  (\psi^+_+\highchi^+_+)\right] +c.c\cr
  &\quad i\kappa\left[ -\alpha_3\psi^+_+\highchi^+_+D^+_+
  \highchi^-_+ -(\alpha_1-i\alpha_2)\psi^+_+\highchi^+_+
  D^+_+\highchi^-_+\right] +c.c\cr
  &                                                  &(7.34)\cr}$$
$$\psi_{+1}D^2_+\psi_{+3}=2\alpha_1\bar\alpha_3\psi^+_+\partial_{++}
  \highchi^-_++2\alpha_1\alpha_3\psi^+_+\partial_{++}\highchi^+_+ +c.c

                \eqno(7.35)$$
$$\eqalignno{
  -\psi_{+1}D_+(\psi_{+2}\psi_{+3}) &= i\alpha_2\bar\alpha_3\left[\alpha_1
  \psi^+_+\highchi^-_+D^-_+\psi^+_+-\bar\alpha_1\psi^-_+\psi^+_+
  D^+_+\highchi^-_+\right] +c.c\cr
  &\quad -i\alpha_1\bar\alpha_2\alpha_3\psi^+_+\highchi^+_+D^+_+\psi^-_+
  +c.c\cr
  &\quad -i\alpha_1\bar\alpha_2\bar\alpha_3\psi^+_+D^+_+(\psi^-_+
  \highchi^-_+) +c.c\cr
  &\quad + i\alpha_2\alpha_3\alpha_1\psi^+_+D^-_+(\psi^+_+\highchi^+_+)
  + c.c
                 &(7.36)\cr}$$
$$\psi_{+1}\psi_{+2}\psi_{+3}\psi_{+4}=-\kappa^2\psi^+_+\psi^-_+\highchi^+_+
  \highchi^-_+\ .
         \eqno(7.37)$$
Disregarding the $U(1)$ charged terms $L^+_{++++}$ and $L^{++}_{++++}$
as they break $N=2$ supersymmetry down to $N=1$, we get the $N=2$
superfield realization of the conformal spin 2 current we are looking for
$$\eqalignno{
  T_{++++} &= \psi^+_+\psi^-_+\highchi^+_+\highchi^-_+ +
  a_0\partial_{++}D^-_+\highchi^+_+ +\bar a_0\partial_{++}
  D^+_+\highchi^-_+\cr
  &\quad -{2i\over\kappa}\ \partial_{++} (\highchi^+_+
  \highchi^-_+)+2i{\alpha_2\bar\alpha_3\over\kappa^2}\
  \partial_{++}(\psi^+_+\highchi^-_+) -2i{\bar\alpha_2\alpha_3
  \over\kappa^2}\ \partial_{++} (\psi^-_+\highchi^+_+)\cr
  &\cr
  &\quad +\psi^+_+\psi^-_+\left[ b_0D^-_+\highchi^+_+-
  \bar b_0D^+_+\highchi^-_+\right] +\psi^+_+\highchi^-_+
  \left[ c_0D^-_+\highchi^+_+ +d_0D^-_+\psi^+_+\right]\cr
  &\quad +\psi^-_+\chi^+_+\left[\bar c_0D^+_+\highchi^-_+ +
  \bar d_0D^+_+\psi^-_+\right] + \psi^+_+\left[
  \bar b_1\partial_{++}\highchi^-_++\bar b_2D^+_+(\psi^-_+
  \highchi^-_+)\right]\cr
  &\quad +\psi^-_+\left[ b_1\partial_{++}\highchi^+_+ +
  b_2D^-_+(\psi^+_+\highchi^+_+)\right]\cr
  &\quad +d_1\highchi^-_+D^-_+(\psi^+_+\highchi^+_+)+
  \bar d_1\highchi^+_+D^+_+(\psi^-_+\highchi^-_+)\cr
  &\quad + D^-_+\psi^+_+\left[ c_1D^+_+\highchi^-_++f_1D^-_+
  \highchi^+_+\right] +D^+_+\psi^-_+\left[\bar e_1D^-_+
  \highchi^+_++\bar f_1D^+_+\highchi^-_+\right]\ ,
&(7.38)\cr}$$
where
$$\eqalignno{
  a_0 &= 2(\alpha_3-i\alpha_4)/\kappa^2,\ \ \  b_0=
  {-(\alpha_4+i\alpha_3)\kappa -i\alpha_1\bar\alpha_2
  \alpha_3\over\kappa^2}\cr
  c_0 &= {\alpha_2+i\alpha_1\over\kappa};\ \ \  d_0=-i\alpha_1
  \alpha_2\bar\alpha_3/\kappa^2\cr
  \bar b_1 &= {2(\alpha_1-i\alpha_2)[(\alpha_3-i\alpha_4)+(
  \bar\alpha_3+i\bar\alpha_4)]-2\alpha_1\bar\alpha_3\over
  \kappa^2}\cr
  \bar b_2 &={i\alpha_2\kappa +i\alpha_1\bar\alpha_2
  \bar\alpha_3\over\kappa^2}\cr
  d_1 &= -{\alpha_2+i\alpha_1\over\kappa}\cr
  e_1 &={-(\alpha_1-i\alpha_2)(\bar\alpha_2+i\bar\alpha_4)\over
  \kappa^2}\cr
  f_1 &= {-(\alpha_1-i\alpha_2)(\alpha_3-i\alpha_4)\over
  \kappa^2}\ .
        &(7.39)\cr}$$
It is worthwhile to mention that the $N=2$ superfield realization of
the conformal spin 2 current given above may be compared to that
obtained recently in Ref.[16] by using another technique. However, we
have noted that within our method based on the $\hat A(2\vert 2)^{(1)}$
Miura transformation, the term of the form $(D^-_+\highchi^+_+)
\highchi^-_+\highchi^+_+$ appearing in the realization of [26] does not
figure in Eq.(7.38). Note that all terms of the conformal spin 2 current
$T_{++++}$ are at most quadratic in the superfields and $\psi^\pm_+$
and $\highchi^\pm_+$ and their derivatives. This is a property of
the $\hat A(2\vert 2)^{(1)}$ Miura transformation as may be seen from
Eqs.(5.8) and (7.9).

\SUBSECTION{The $U_5$ supercurrent}

As for the $U_3$ supercurrent, the hermitian $U_5$ current is given by
the real part of a $N=2$ complex conformal spin ${5\over 2}$ current
$\Gamma^+_{5/2}$. The chirality property of the $\psi^\pm_+$ and
$\highchi^\pm_+\ N=2$ superfields shows that $U_5$ exhibits the
following $U(1)$ charge decomposition
$$U_5=\kappa^2\left(\Gamma^+_{5/2}+\Gamma^-_{5/2}\right)\ .

           \eqno(7.40)$$

Using Eq.(7.10) and Eqs.(7.7)--(7.14), one finds after some algebra
$$\eqalignno{
  \Gamma^+_{5/2} &= \psi^+_+\psi^-_+\highchi^+_+D^+_+
  \highchi^-_+ +{2\over\kappa}\ \alpha_4\ \psi^+_+\psi^-_+
  \partial_{++}\highchi^+_+\cr
  &\quad +\psi^+_+\Biggl[{2i\alpha_1\over\kappa}\
  \partial_{++}(\highchi^+_+\highchi^-_+)+{2\alpha_1\alpha_3\over
  \kappa^2}\ \partial_{++}D^-_+\highchi^+_+ +
  {2\alpha_1\bar\alpha_3\over\kappa^2}\ \partial_{++}
  D^+_+\highchi^-_+\cr
  &\qquad -{2i\over\kappa}\ (\alpha_1-i\alpha_2)\left[
  D^+_+(\highchi^-_+D^-_+\highchi^+_+)-D^-_+(\highchi^+_+D^+_+
  \highchi^-_+)\right] + {2i(\alpha_1-i\alpha_2)\over\kappa^2}\
  \bar\alpha_4\cdot\partial_{++}D^+_+\highchi^-_+\Biggr]\cr
  &\quad -{2i\over\kappa}\left[ (\alpha_1-i\alpha_2)D^-_+
  \psi^+_++(\bar\alpha_1+i\bar\alpha_2)D^+_+\psi^-_+\right]
  \highchi^+_+D^+_+\highchi^-_+ + {2i\over\kappa^2}\
  (\bar\alpha_1+\bar\alpha_2)\bar\alpha_4 D^+_+\psi^-_+
  \partial_{++}\highchi^+_+\cr
  &\quad -{2i\over\kappa^2}\ (\alpha_1-i\alpha_2)\alpha_4\
  D^+_+\psi^-_+\partial_{++}\highchi^+_+ +
  {4i\alpha_4\over\kappa^2}\ \partial^2_{++}\highchi^+_+ -
  {2i\over\kappa}\ \partial_{++} (\highchi^+_+D^+_+\highchi^-_+)\ .

                &(7.41)\cr}$$

\SECTION{$N=2$ BOUSSINESQ EQUATION}

The Lie (super) algebraic approach provides a systematic method
of constructing generalized KdV type equations and their supersymmetric
extensions [3, 4, 5]. In the bosonic case, one associates with the affine
Lie algebria $\hat A_{n-1}^{(1)}$ a scalar Lax operator of $n$--th order
$P(\partial )$. The latter is a polynom of the differential operator
$\partial_z$ given by; see Eqs.(4.27) and (4.28)
$$L_b=\partial^n+\sum^n_{b=2}\ U_k\ \partial^{n-k}\ ,          \eqno(8.1)$$
where we have set
$$\eqalign{
  P(\partial ) &= L_b\cr
  \partial &= \partial_{++} =\partial_z\ .\cr}
\eqno(8.2)$$
Then the following hierarchy of time evolution equations
$${\partial L_b\over\partial t_k} =\left[ (L_b)^{k/n}_+, L_b
  \right]\ ,
            \eqno(8.3)$$
gives the generalized KdV hierarchy of $A_{n-1}$ type. Here, the
subscript $+$ means the differential part of the pseudo operator
$(L_b)^{k/n}$. It should not be confused with the convention notations
of the previous sections.

In the supersymmetric case, a particularly interesting class of hierarchy
of super Lax equations based on the affine Lie superalgebras
$\hat A(n-1\vert n-1)^{(1)}$ can be constructed. There, the scalar Lax
superoperator $P({\cal D}^{(1)})$ Eqs.(4.27)--(4.28) is of $2n$--ith order
and reads as
$$L_s =D^{2n}+\sum^{n-1}_{k=1}\ \left( U_{2k}\ D^{2n-2k} +
  U_{2k+1}\ D^{2n-2k-1}\right)\ ,
\eqno(8.4)$$
where we have set
$$\eqalign{
  P({\cal D}^{(1)}) &= L_s\cr
  D &= D_+ =D_\theta\ .\cr}
 \eqno(8.5)$$
The generalized supersymmetric KdV hierarchy of type $A(n-1\vert n-1)$
reads as [5, 15]
$${\partial L_s\over\partial t_{2k}} =\left[ \left( L^{{2k\over 2n}}_2
  \right)_{>0},L_s\right] : k\not= 3\ell\ .
\eqno(8.6)$$
The subscript $>0$ means taking the strictly positive differential
operator part of the super pseudodifferential operator $L^{2k/2n}_s$ as
time evolution operator. This prescription is necessary for incorporating
in the game the conformal spin one current $U_2$.

The aim of this section is to write down the $N=2$ supersymmetric
Boussinesq equation using $N=2$ superfield language. We also give
a generalization of this equation involving non hermitian supercurrents.

Setting $n=3$ and writing Eq.(8.4) as
$$L=D^6-UD^4-VD^3-RD^2-SD\ ,
\eqno(8.7)$$
the generalized $N=2$ super KdV hierarchy of $A(2\vert 2)$ type reads as
$${\partial L\over\partial t_{2k}} =\left[\left( L^{{2k\over 6}}
  \right)_{>0},L\right]\ .
      \eqno(8.8)$$
The first non trivial time evolutions are obtained by taking $k=2$.
In this case
$$\left( L^{{4\over 6}}\right)_{>0} =D^4-{2\over 3}\ UD^2-{2\over 3}\ VD\ .

                  \eqno(8.9)$$

Following Ref.[27], the time evolution equations resulting from Eq.(8.8)
are given by
$$\eqalignno{
  {\partial U\over\partial t_4} &= -D^4U+UD^2U+2D^2R &(8.10)\cr
  &\cr
  {\partial V\over\partial t_4} &= -D^4V+UD^2V+VD^2U+2D^2S &(8.11)\cr
  &\cr
  {\partial R\over\partial t_4} &= D^4R+{2\over 3}\left[ -D^6U+UD^4U-
  UD^2R+RD^2U\right]\cr
  &\quad +{2\over 3}\left[ VD^3U-VD^2V-VDR+2VS+SDU\right] &(8.12)\cr
  &\cr
  {\partial S\over\partial t_4} &= D^4S+{2\over 3}\left[
  -D^6V-UD^2S+UD^4V+VD^3V+RD^2V-VDS+SDV\right]\ .        &(8.13)\cr}$$
Before going ahead note that Eqs.(8.11) and (8.13) reduce respectively,
to Eqs.(8.10) and (8.12) if the following choice is made
$$\eqalign{
  V &= DU\cr
  S &= DR\ .\cr}
     \eqno(8.14)$$
Observe that the last term of Eq.(8.12) vanishes identically once the
representation (8.14) is used.

Using the $U(1)$ charge decomposition of the supercurrents appearing
in Eq.(8.1), namely
$$\eqalign{
  U &=\kappa\ J+J^{++}+J^{--}\cr
  V &=i\kappa (F^+-F^-)\cr
  R &=\kappa^2T+\kappa L^{++}+\kappa L^{--}\cr
  S &=\kappa^2(\Gamma^++\Gamma^-)\ ,\cr}                  \eqno(8.15)$$
Eqs.(8.10)--(8.13) may be rewritten in terms of the $U(1)$ blocks
$J,T, F^\pm ,\Gamma^\pm ,J^{\pm\pm}$ and $L^{\pm\pm}$. In the total
there are ten coupled $N=2$ superfield equations classified as follows:
1)\ Two real equations giving the time evolutions of the $J$ and $T$
superfields. These equations describe the $N=2$ supersymmetric
extension of the Boussinesq equation. 2)\ Two complex equations giving
the time evolution of the $N=2$ superfields $J^{++}$ and $L^{++}$. The
equations corresponding to $J^{--}$ and $L^{--}$ are obtained by taking
the complex conjugate of the previous ones. 3) Two complex equations
describing the time evolution of $F^+$ and $\Gamma^+$. Their complex
conjugates imply those of $F^-$ and $\Gamma^-$. We give hereafter the time
evolution equations of $J, J^{++}, F^+, T, L^{++}$ and $\Gamma^+$ as well
as the consistency conditions
$$\eqalignno{
  \dot J &= 2\kappa (J\partial J+2\partial T)-4\partial^2J+{2\over \kappa}\
  \partial (J^{++}J^{--} )&(8.16)\cr
  &\cr
  \dot J^{++} &=2\kappa\partial\left[ (JJ^{++})+2L^{++}\right] +
  4\partial^2J^{++} &(8.17)\cr
  &\cr
  \dot F^+ &= 2\kappa\left[ \partial (JF^+)-\partial (J^{++}F^-)
  -2i\partial\Gamma^+\right] -4\partial^2F^+ &(8.18)\cr
  &\cr
  \dot T &= {4\over 3}\kappa\left[\partial (JT)+i(F^+\Gamma^--F^-\Gamma^+)+
  {i\over 2}(F^+D^-T-F^-D^+T)+{1\over 2}(\Gamma^+D^-J+\Gamma^-
  D^+J)\right]\cr
  &\quad +4\partial^2T+{4\over 3}\Biggl[ J\partial^2J+F^+\partial F^-+
  F^-\partial F^+-{i\over 2} F^+D^+L^{--}+{i\over 2}F^- D^-L^{++}\cr
  &\qquad +{1\over 2}(\Gamma^+D^+J^{--}+\Gamma^-D^-J^{++})+
  4i(F^+\partial D^-J-F^-\partial D^+J)\Biggr]\cr
  &\quad +{-4\over 3\kappa}\left[ J^{++}\partial L^{--}-L^{--}\partial J^{++})
+
  (J^{--}\partial L^{++}-L^{++}\partial J^{--})- 4i(F^+\partial D^+J^{--}-
  F^-\partial D^-J^{++})\right]\cr
  &\quad +{8\over 3\kappa^2} \left[J^{++}\partial^2J^{--}+J^{--}\partial^2
  J^{++}\right]   &(8.19)\cr}$$
$$\eqalignno{
  \dot L^{++} &= {4i\over 3}\ \kappa^2\ F^+\Gamma^+\cr
  &\quad +{4\over 3}\kappa\Biggl[ (L^{++}\partial J-J
  \partial L^{++})+(T\partial
  J^{++}-J^{++}\partial T)-F^+\partial F^+\cr
  &\qquad +4i(F^+\partial D^+J+F^+\partial D^-J^{++}) -
  {i\over 2} (F^+D^+T-F^+D^-L^{++})\cr
  &\qquad +2(\Gamma^+D^+J+\Gamma^+D^-J^{++})\Biggr]\cr
  &\quad +4\partial^2L^{++} +{8\over 3} [J\partial^2J^{++}+J^{++}
  \partial^2J]-{16\over 3\kappa}\ \partial^3J^{++} &(8.20)\cr
  \dot\Gamma^+ &= {2i\kappa\over 3}\Biggl[
  (\Gamma^+D^-F^+-\Gamma^+D^+F^-)
  -(F^-D^+\Gamma^+-F^+D^+\Gamma^-)\cr
  &\qquad +(\Gamma^-D^+F^+-F^+D^-\Gamma^+) +2T\partial_z
  F^++4iJ\partial\Gamma^+\Biggr]\cr
  &\quad +4\partial^2\Gamma^+-{4\over 3} i\Biggl[ L^{++}
  \partial_zF^- -2i(F^+\partial D^-F^+-F^-\partial D^+F^+)\cr
  &\qquad +2i (F^+\partial D^+F^-+J^{++}\partial\Gamma^-_
  -2J\partial^2F^+\Biggr]\cr
  &\quad -{8i\over 3\kappa}\left[ 2\partial^3F^++J^{++}\partial^2F^+\right]\ .

                 &(8.21)\cr}$$
 In writing Eqs.(8.16)--(8.21), we have used the convention notation
 ${\partial\phi\over\partial t_4}=\dot\phi$. The consistency conditions
 read, however, as
 $$\eqalignno{
   \partial (J^{++})^2 &= 0 &(8.22)\cr
   \partial (J^{++}F^+) &= 0 &(8.23)\cr}$$
 $$\eqalignno{
   &\kappa^2[\Gamma^+D^+J^{++}-iF^+F^+L^{++}] +2\kappa [L^{++}\partial J^{++}-
   J^{++}\partial L^{++}+8iF^+\partial D^+J^{++}]\cr
   &\quad +4J^{++}\partial^2J^{++} =0 &(8.24)\cr
   &\cr
   &\kappa^2\left[ i(\Gamma^+D^+F^+-F^+D^-\Gamma^+)\right] -\kappa
   \left[ 4(J^{++}\partial\Gamma^++F^+\partial D^+F^+)+2iL^{++}
   \partial F^+\right]\cr
   &\quad +4iJ^{++}\partial^2F^+ =0\ .    &(8.25)\cr}$$

 Finally, note that Eqs.(8.16)--(8.21) describe a generalized version of the
 following time evolution relations
 $$\eqalignno{
   \dot J &= -4\partial^2J+2\kappa (J\partial J+2\partial T) &(8.26)\cr
   &\cr
   \dot T &= -4\partial^2T +{8\over 3}\ J\partial^2J\cr
   &\quad +{8\kappa\over 3}\left[ T\partial J-{1\over 2}\ J\partial
   T-2\partial^3J\right] &(8.27)\cr}$$
 where $J$ and $T$ are the $N=2$ conformal spin one and two
 supercurrents. These equations which are obtained from Eqs.(8.10),
 (8.12) and (8.14) by setting $U=\kappa J$ and $R=\kappa^2T$ give
 a $N=2$ supersymmetric extension of the Boussinesq equation. They
 have the same structure as those derived in Ref.[26].

 \SECTION{CONCLUSION}

 In this paper we introduced the concept of conformal spin gradation
 of the untwisted affine Lie superalgebra $\hat A(n-1\vert n-1)^{(1)}$ to
 study the $W\hat A(n-1\vert n-1)^{(1)}$ Miura transformation and its
 implications in $N=2$ affine integrable theories. Our results may be
 summarized as follows.

 1.\quad We developed the concept of conformal spin gradation of
 $\hat A(n-1\vert n-1)^{(1)}$ according to which the fundamental
 generators $E_{\pm\alpha_a}$ associated with the purely fermionic
 simple root system $\{\alpha_a, a=(i,2n)\}$ of $\hat A(n-1\vert n-1)^{(1)}$
 have the same conformal properties as two dimensional space--time
 Weyl spinors. All the other $\hat A(n-1\vert n-1)^{(1)}$ objects are built out
 of the $E_{\pm\alpha_a}$'s and behave as higher two dimensional
 space--time Lorentz representations. Moreover, as $A(n-1\vert n-1)$
 is realized in terms of $2n\times 2n$ matrices, we noted that conformal
 spin gradation of $\hat A(n-1\vert n-1)^{(1)}$ is carried by the vector
 basis $\vert a>$ of the $SL(n\vert n)$ vector representation space
 $V_{2n}$. We gave the generic formula of their conformal spin weights
 Eq.(3.1). The latter allow the determination of any $\hat A(n-1\vert n-1)
 ^{(1)}$ quantity and, too particularly, the algebra generators Eqs.(3.2),
 (3.3) and (3.5), (3.6). Thus any $2n\times 2n$ matrix element $M$ of
 $\hat A(n-1\vert n-1)^{(1)}$ may be written as
 $$M=\sum^{2n}_{a,b=1}\ m_{ab}\vert a><b\vert\ .               \eqno(9.1)$$
 The conformal spin wieght of the matrix element $m_{ab}$ depends on
 both the CSW of $M$ and the gradations of $\vert a>$ and $\vert b>$ as
 shown below
 $${\rm CSW}(m_{ab})={\rm CSW}(M)-(b-a)\ ,                       \eqno(9.2)$$
 if both $a$ and $b$ belong to the set $\{ 1,\dots ,n\}$ or
 $\{ n+1,\dots ,2n\}$. In case where $a$ belongs to $\{ 1,\dots ,n\}$ and
 $b$ belongs to $\{ n+1,\dots ,2n\}$ or inversely, Eq.(9.2) is replaced by
 $${\rm CSW}(m_{ab})={\rm CSW}(M) -(b-a)\mp \left( n-{1\over 2}
   \right)\ .
          \eqno(9.3)$$
 Note that Eqs.(9.2) and (9.3) obey the property
 $${\rm CSW}(m_{ab})+{\rm CSW}(m_{ab})=2{\rm CSW}(M)\ .   \eqno(9.4)$$
 The so--called diagonal and canonical gauges of the superfield $Q_+$,
 involved in the supersymmetric Lax operator Eqs.(4.14) and (4.33), are
 really the best illustration of the conformal spin gradation idea. Indeed
 in the diagonal gauge Eq.(4.19), $Q_+$ reads as
 $$Q^{(0)}_+=\sum^{2n}_{i=1}\ (Q_+)_{ii}\ \vert i><i\vert\ ,    \eqno(9.5)$$
 where the diagonal elements $(Q_+)_{ii}$ are two dimensional Weyl
 spinors and the $\vert i><i\vert$ projectors are conformal scalars. Using
 the Kac notation $SL(n\vert n)=SL_{\bar 0}(n\vert n)\oplus SL_{\bar 1}
 (n\vert n)$, one sees that $Q^{(0)}_+$ has the structural form
 $$Q^{(0)}_+\sim [{\rm space-time\ spinors}]\otimes
   SL_{\bar 0}(n\vert n)\ ,
  \eqno(9.6)$$
 where $SL_{\bar 0}(n\vert n)$ is the even part of the superalgebra
 $SL(n\vert n)$. In the canonical gauge however, $Q_+$ reads as
 $$Q^{(1)}_+=\sum^{n=1}_{k=1}\ \left[ U_{2k+1}\vert k><1\vert +
   U_{2k}\vert k+n><1\vert\right]\ ,
\eqno(9.7)$$
 where $U_{2k+1}$ and $U_{2k}$ are respectively space--time conformal
 spinors and bosons. The step operators $\vert k><1\vert$ and $\vert k+n
 ><1\vert$ belong to $SL_{\bar 0}(n\vert n)$ and $SL_{\bar 1}(n\vert n)$
 respectively. Thus the structural form of $Q^{(1)}_+$ is
 $$Q^{(1)}_+\sim ({\rm space-time\ spinors})\otimes SL_{\bar 0}
   (n\vert n) +({\rm space-time\ bosons})\otimes SL_{\bar 1}
   (n\vert n)\ .
      \eqno(9.8)$$
 As $Q^{(0)}_+$ and $Q^{(1)}_+$ are gauge equivalent, one concludes that
 elements of $SL_{\bar 0}(n\vert n)$ behave as space--time bosons
 whereas elements of $SL_{\bar 1}(n\vert n)$ behave as space--time
 fermions.

 We studied also the properties of $\Lambda_+$, the sum of conformal
 spin ${1\over 2}$ generators. We found that they obey a set of remarkable
 properties among which we have: 1) Decomposability showing that
 $\Lambda_+$ splits in two irreducible and nilpotent parts $\lambda^\pm_+$
 Eqs.(1.2). This feature together the reducibility of the $\Lambda^+_+
 \Lambda^-_+$ product namely
 $$2\Lambda^+_+\cdot\Lambda^-_+ =\Lambda_{++}+\Sigma_{++}\ ,

               \eqno(9.9)$$
 where $\Lambda_{++}$ and $\Sigma_{++}$ are given by Eqs.(3.12)--(3.14),
 play a crucial role in the setting of a manifestly $N=2\ U(1)$ Lax formalism.

 2.\quad We analysed the $W\hat A(n-1\vert n-1)^{(1)}$ Miura
 transformation leading to $N=2$ superfield realization of the $W_s$
 supercurrents with $1\leq s\leq n-1$. First, we reviewed the
 $\hat A(n-1\vert n-1)^{(1)}$ Lax formalism using $N=1$ superfield
 techniques. Then, we showed that the $2(n-1)$ obtained $N=1$
 superfield currents, $U_k; 2\leq k\leq 2n-1$, can be cast into $(n-1)\ N=2$
 real multiplets $J_k=U_{2k}\oplus U_{2k+1}, 1\leq k\leq n-1$.
 This structure shows that the higher spin $N=2$ supercurrent content of
 $\hat A(n-1\vert n-1)$ is analogous to the bosonic $\hat A_{n-1}$ case.
 There, the higher conformal spin $N=0$ current content is given by
 $\omega_k$ with $2\leq k\leq n$. $W_1$ is forbidden by the traceless
 condition of the $\widehat{SL}(n)$ \ $n\times n$ matrix representation.
 Using these features and those of $\Lambda_+$ Eqs.(3.10)--(3.12), we
 develop a manifestly $N=2$ supersymmetric Lax formalism leading in
 turns to a manifestly $N=2$ supersymmetric $W\hat A(n-1\vert n-1)$
 Miura transformation, namely
 $$\nabla^n_{++} +\sum^{n-1}_1\ T_k\ \nabla^{n-k}_{++} =
   \left[ \mathop{\Pi}\limits^{n-1}_{j=1}\ (\nabla_{++} -J_{++})_j
   \right]\nabla_{++}\ ,
    \eqno(9.10)$$
 where $\nabla_{++}=[D^+_+,D^-_+]$. $J_{++}$ is a composite conformal
 spin one object given by Eqs.(4.50) and (4.51). Eq.(9.10) gives directly
 the $N=2$ superfield realization of the higher spin $N=2$
 supercurrents. The $\hat A(1\vert 1)^{(1)}$ and $\hat A(2\vert 2)^{(1)}$
 leading examples were discussed, see Eqs.(4.52). However, this approach,
 which seems the natural way to deal with $N=2$ higher conformal
 spin symmetries, is obtained by imposing constraints on the $N=2$
 supersymmetric Lax operators Eqs.(4.33) and (4.35), (4.36). A detailed
 analysis of these constraints is lacking.

 Moreover, we developed a new method of constructing $N=2$ superfield
 realization of the $N=2$ super $W$--currents from the $N=1$ superfield
 formulation of the $W\hat A(n-1\vert n-1)^{(1)}$ Miura transformation.
 The idea may be summarized as follows:

 \noindent  i) First, break the $N=2$
 supersymmetric Lax formalism down to $N=1$ as in Eqs.(5.1)--(5.7).
 Thus the resulting $N=1$ Lax analysis can be viewed as the real part
 truncation of the $N=2$ one. Formally, this truncation may be read as
 $$(N=1)\sim (N=2)+\overline{(N=2)}\ .
\eqno(9.11)$$
 ii) Take the results of the $N=1$ superfield language of the $W\hat A(n-1
 \vert n-1)$ analysis of Subsection 4.2 and replace all $N=1$ quantities
 by their analogue of type Eq.(9.11).

 \noindent  iii) Use the $U(1)$ charge of the
 $N=2$ superalgebra to classify the resulting spectre in $U(1)$ blocks.
 We found that the $U_k, 2\leq k\leq 2n-1$ supercurrents decompose as
 $$\eqalignno{
   U_{2k} &= \kappa J^0_k+J^{++}_k+J^{--}_k;\quad\
   1\leq k\leq n-1 &(9.12)\cr
   U_{2k+1} &=\kappa\left(\Gamma^+_{k+{1\over 2}} +
   \Gamma^-_{k+{1\over 2}}\right):\quad 1\leq k\leq n-1\ . &(9.13)\cr}$$
 Higher $U(1)$ charge representation are forbidden by $N=2$ chirality.
 $J^0_k$ is the only hermitian block. Furthermore, given an integer $n$,
 there are $(n-1)\ N=2$ real supercurrents of conformal spin
 $k$;\hfil\break
 $1\leq k
 \leq n-1$. These features together with $U(1)$ covariance ensure that
 the $J^0_k$'s are indeed the $N=2$ $W_k$--supercurrents predicted by
 the general arguments of Subsection 4.2, Eqs.(4.23)--(4.25). The other
 blocks $\Gamma^\pm_{k+{1\over 2}}$ and $J^{\pm\pm}_k$ are
 $N=2$ complex
 supercurrents and deal with non unitary theories.

 To illustrate this method, we studied in detail the $\hat A(1\vert 1)^{(1)}$
 and $\hat A(2\vert 2)^{(1)}$ theories. For the first case, we obtained the
 general form of the spin one $N=2$ supercurrent recovering all the known
 realizations. For the $\hat A(2\vert 2)^{(1)}$ case, we gave the full
 analysis of the five involved supercurrent blocks. For the interesting
 hermitian block $(J=J^0_1,T=J^0_2)$ an unexpected result is obtained.
 We found a new Feigin Fuchs type extension built out of $\psi^\pm$
 and $\highchi^\mp$ Eq.(7.25). It can be put on equal footing with
 the usual Feigin Fuchs extension based on $D^\pm$ and $\psi^\mp$.
 Note that $\psi^\pm$ and $\highchi^\pm$ are the two $N=2$ superfields
 involved in the $W\hat A(2\vert 2)^{(1)}$ Miura transformation. Note
 also that the new Feigin Fuchs extension $\psi^-\highchi^+$ was not
 predicted by the $N=2$ manifestly $W\hat A(n-1\vert n-1)$ Miura
 transformation. This fact let understand that the constraints we took
 Eqs.(4.35) and (4.36) are very restrictive. For the spin two $N=2$
 supercurrent $T$ containing the bosonic spin three current, things
 follow in a straightforward way. However, it should be noted that our
 result compared with a recent $N=2$ superfield realization of $T$ [26]
 differs by a term of the form $\highchi^+_+\highchi^-_+D^+_+\highchi^-_+$.
 The $W\hat A(2\vert 2)^{(1)}$ method used in here allows only terms
 at most quadratic in the superfields and their derivatives.

 3.\quad We gave the $N=2$ supersymmetric extension of the
 Boussinesq equation by using the $N=1$ superfield $\hat A(2\vert
 2)^{(1)}$ Lie superalgebraic analysis conjugated with the $U(1)$
 charge decomposition method. A generalization of these $N=2$
 super equations involving complex supercurrents were also obtained.

 \centerline{APPENDIX A}
 \bigskip

 Here we give the derivation of Eqs.(7.7)--(7.10). They are obtained
 after four steps of calculations.

 1.\quad Calculate the quantity $A$
 $$\eqalignno{
   A &= (D-\xi_5)(D^2-\xi_3D)\cr
   &= D^3+a_2D^2+a_1D   &(A.1)\cr}$$
 where
 $$\eqalignno{
   a_2 &= \xi_3-\xi_5=-\psi_3 &(A.2)\cr
   a_1 &=\xi_5\xi_3-D\xi_3=\psi_3\psi_4-D\psi_4\ ,   &(A.3)\cr}$$
 where Eqs.(7.5) have been used.

 2.\quad Calculate $B=(D-\xi_2)A$
 $$B=D^4+b_3D^3+b_2D^2+b_1D
\eqno(A.4)$$
 with
 $$\eqalignno{
   b_3 &= -(a_2+\xi_2)=-(\xi_2+\xi_3-\xi_5)=-\psi_2 &(A.5)\cr
   &\cr
   b_2 &= a_1+Da_2-\xi_2a_2\cr
   &=-D\xi_5+\xi_5\xi_3-\xi_2\xi_3+\xi_2\xi_5 &(A.6)\cr
   &=-D(\psi_3+\psi_4)+\psi_2\psi_3+\psi_3\psi_4\cr
   &\cr
   b_1 &= Da_1-\xi_2a_1\cr
   &=-D^2\xi_3+D(\xi_5\xi_3) &(A.7)\cr
   &=-D^2\psi_4+D(\psi_3\psi_4)\ .  \cr}$$

 3.\quad Calculate $C=(D-\xi_4)B$
 $$c = D^5+c_4D^4+c_3D^3+c_2D^2+c_1D                             \eqno(A.8)$$
using Eqs.(A.3) and (A.4), we get
\vfill\eject
 $$\eqalignno{
   c_4 &= -(b_3+\xi_4)=(\xi_2+\xi_3-\xi_5-\xi_4)=-\psi_1 &(A.9)\cr
   &\cr
   c_3 &= (b_2+Db_3-\xi_4b_3)\cr
   &=-D(\xi_2+\xi_3)+\xi_1\xi_4-\xi_2\xi_3+\xi_5\xi_3+\xi_2\xi_5\cr
   &= -D(\psi_2+\psi_3+\psi_4)+\psi_2\psi_3+\psi_3\psi_4
   +\psi_1\psi_2 &(A.10)\cr
   &\cr
   c_2 &= Db_2-\xi_4b_2-b_1\cr
   &=-D^2(\xi_5-\xi_3)-D(\xi_2\xi_3)+D(\xi_2\xi_5)+\xi_4D\xi_5+
   \xi_4\xi_2\xi_3 -\xi_4\xi_2\xi_5-\xi_4\xi_5\xi_3\cr
   &=-D^2\psi_3+D(\psi_2\psi_3)-(\psi_1+\psi_2)D(\psi_3+\psi_4)-
   \psi_1\psi_2\psi_3-\psi_1\psi_3\psi_4-\psi_2\psi_3\psi_4\cr
   &     &(A.11)\cr
   c_1 &= Db_1-\xi_4b_1\cr
   &=-D^3\xi_3+D^2(\xi_5\xi_3)+\xi_4D^2\xi_3-\xi_4D(\xi_5\xi_3)\cr
   &=-D^3\psi_4+D^2(\psi_3\psi_4)-(\psi_1+\psi_2)D(\psi_3\psi_4 )
   +(\psi_1+\psi_2)D^2\psi_4\ . &(A.12)\cr}$$

 4.\quad Finally, we calculate the desired quantity $R$
 $$\eqalignno{
   R &= (D-\xi_1)(D-\xi_4)(D-\xi_2)(D-\xi_5)(D-\xi_3)D\cr
   &= (D-\xi_1)\cdot C\ .
&(A.13)\cr}$$
 Expanding $R$ as
 $$R = D^6+r_5D^5+r_4D^4+r_3D^3+r_2D^2+r_1D                   \eqno(A.14)$$
 and using Eqs.(A.8)--(A.12), we find
 $$\eqalignno{
   U_1 &= r_5=-(c_4+\xi_5)=0 &(A.15)\cr
   U_2 &=r_4=(c_3+Dc_4-\xi_1c_4) &(A.16)\cr
   U_3 &=r_3=(Dc_3-\xi_1c_3-c_2) &(A.17)\cr
   U_4 &= r_2=(Dc_2-\xi_1c_2+c_1) &(A.18)\cr
   U_5 &= r_1=(Dc_1-\xi_1c_1)        &(A.19)\cr}$$
 where the $U_i$'s are given by Eqs.(7.7)--(7.10).
 \vfill\eject

 \centerline{APPENDIX B}
 \bigskip

 To calculate the $N=2$ superfield realization of the complex
 supercurrents $K^+_{3/2}$ and $\Gamma^+_{5/2}$, Eqs.(7.25) and (7.41),
 the following formulas are useful.

 \noindent I.\quad The $K^\pm_{3/2}$ case

 Recall that
 $$U_3-i\kappa (K^+_{3/1}-K^-_{3/2})\ .
\eqno(B.1)$$

 The identification of $K^\pm_{3/2}$ is achieved by using Eq.(7.8) and
 the identities
 $$\eqalignno{
   &1)\ \ D^2(\psi_2+\psi_4) = (i\alpha_2D^+D^-\psi^++i\alpha_4
   D^+D^-\highchi^+)+c.c &(B.2)\cr
   &\cr
   &2)\ \ D(\psi_1\ \psi_2) = i\kappa [(D^-\psi^+)\psi^- -
   \psi^+D^+\psi^-] &(B.3)\cr
   &\cr
   &3)\ \ D(\psi_3\ \psi_4) =i\kappa [D^-(\highchi^+\highchi^-) +
   D^+(\highchi^+\highchi^-)] &(B.4)\cr
   &\cr
   &4)\ \ \psi_1D\psi_2 = i\alpha_1\psi^+[\bar\alpha_2D^+\psi^- -
   \alpha_2D^-\psi^+] +c.c &(B.5)\cr
   &\cr
   &5)\ -\psi_2D\psi_3 = i\alpha_2\psi^+[\bar\alpha_3D^+
   \highchi^- +\alpha_3D^-\highchi^+] +c.c &(B.6)\cr
   &\cr
   &6)\ -\psi_2D\psi_4 =\alpha_2\psi^+(\alpha_4D^-\highchi^+ -
   \bar\alpha_4D^+\chi^-) +c.c &(B.7)\cr
   &\cr
   &7)\ \psi_2\psi_3\psi_4 =\alpha_2\kappa\ \psi^+\highchi^+
   \highchi^- +c.c\ .                &(B.8)\cr}$$

 \noindent II.\quad The $\Gamma^\pm_{5/2}$ case.

 Here we have
 $$U_5=\kappa^2(\Gamma^+_{5/2} +\Gamma^-_{5/2})          \eqno(B.9)$$
 where $\Gamma^\pm_{5/2}$ are calculated with the help of
 $$\eqalignno{
   &1)\ -D^4\psi_4=4i\alpha_4\partial^2_z\highchi^+ +c.c &(B.10)\cr
   &\cr
   &2)\ D^3(\psi_3\psi_4)=-2i\kappa\ \partial_z(\highchi^+D^+
   \highchi^-) +c.c &(B.11)\cr
   &\cr
   &3)\ D[(\psi_1+\psi_2)D^2\psi_4]=-2i(\alpha_1-i\alpha_2)
   [\alpha_4D^-(\psi^+\partial_z\highchi^+)\cr
   &\qquad -\bar\alpha_4
   \psi^+\partial_zD^+\highchi^-+\bar\alpha_4D^-\psi^+\partial_z
   \highchi^-] +c.c &(B.12)\cr
   &\cr
   &4)\ -D[(\psi_1+\psi_2)D(\psi_3\psi_4)]= 2i\kappa [(\alpha_1-i
   \alpha_2)D^-\psi^+\cr
   &\qquad +(\bar\alpha_1+i\bar\alpha_2)D^+\psi^-]
   [\highchi^-D^-\highchi^+-\highchi^+D^+\highchi^-]\cr
   &\qquad -2i\kappa [(\alpha_1-i\alpha_2)
   \psi^+ +(\bar\alpha_1+i\bar\alpha_2)\psi^-][D^+\highchi^-
   D^-\highchi^+ -2\highchi^-\partial_z\highchi^+ -c.c] &(B.13)\cr
   &\cr
   &5)\ \psi_1D^3\psi_3 =2\alpha_1\alpha_3\psi^+\partial_zD^-
   \highchi^+ +2\alpha_1\bar\alpha_3\psi^+\partial_zD^+
   \highchi^- +c.c &(B.14)\cr
   &\cr
   &6)\ \psi_1D^2(\psi_3\psi_4) =2i\kappa\alpha_1\psi^+\partial_z
   (\highchi^+\highchi^-) +c.c &(B.15)\cr
   &7)\ (\psi_1\psi_2)D(\psi_3\psi_4) =-\kappa^2\psi^+\psi^-
   [(D^-\highchi^+)\highchi^- -\highchi^+D^+\highchi^-] &(B.16)\cr
   &\cr
   &8)\ -\psi_1\psi_2D^2\psi_4 =2\kappa\psi^+\psi^-(\alpha_4\partial_z
   \highchi^+ -\bar\alpha_4\partial_z\highchi^-)\ . &(B.17)\cr}$$
 \vfill\eject

 \centerline{REFERENCES}
 \bigskip

 \item{[1]}
 A.B. Zamolodchikov, Teor. Math. Fiz. {\bf 65} (1985) 347;\hfil\break
 V.A. Fateev and A.B. Zamolodchikov, Nucl. Phys. {\bf B304} (1988) 348.

 \item{[2]}
 A. Bilal, V.V. Fock and I.I. Kogan, Nucl. Phys. {\bf B359} (1991)
635;\hfil\break
 M. Bershadsky and H. Ooguri, Com. Math. Phys. {\bf 126} (1989) 49;\hfil\break
 G.M. Sotkov, M. Stanishkov and C.J. Zhu, Nucl. Phys. {\bf B356} (1991)
 245;\hfil\break
 G. Sotkov and M. Stanishkov, Nucl. Phys. {\bf B356} (1991) 439.

 \item{[3]}
 V.G. Drinfel'd and V.V. Sokolov, Sov. J. Math. {\bf 30} (1985) 1975 and
 Sov. Math. Dokl. {\bf 23} (1981) 457.

 \item{[4]}
 Yu.I. Manin and A.O. Radul, Com. Math. Phys. {\bf 98} (1985) 65;\hfil\break
 S. Panda and S. Roy, ICTP, Trieste, preprint No.IC/92/70 (1992);\hfil\break
 F. Yu, preprint, University of Utah (1991).

 \item{[5]}
 T. Inami and H. Kanno, Com. Math. Phys. {\bf 136} (1991) 519.

 \item{[6]}
 E.H. Saidi and M. Zakkari, Phys. Lett. {\bf B281} (1992) 67.

 \item{[7]}
 V.A Fateev and S. Lukyanov, Int. J. Mod. Phys. {\bf A3} (1988) 507;\hfil\break
 F. Bais, P. Bouwknegt, M. Surridge and K. Schoutens, Nucl. Phys. {\bf B304}
 (1988) 348;\hfil\break
 L. Romans, Nucl. Phys. {\bf B352} (1991) 829;\hfil\break
 E. Bergshoeff, C.N. Pope, L.J. Romans, E. Sezgin and X. Shen, Phys. Lett.
 {\bf B245} (1990) 447.

 \item{[8]}
 J. Evans and T. Hollowood, Nucl. Phys. {\bf B352} (1991) 529;\hfil\break
 H. Nohara and K. Mohri, Nucl. Phys. {\bf B349} (1991) 529;\hfil\break
 D. Olive and N. Turok, Nucl. Phys. {\bf B257} [FS 14] (1986) 277;\hfil\break
 D. Olive, Lecture Notes in Gauge Theories and Lie Algebras with
 Applications to Integrable Systems, Univ. of Virginia Fall (1882).

 \item{[9]}
 C.M. Hull, Nucl. Phys. {\bf B353} (1991) 707;\hfil\break
 C.M. Hull, Talk given at the Summer Workshop on High Energy Physics,
 ICTP, Trieste (1992) and references therein;\hfil\break
 K. Schoutens, A. Sevrin and P. Van Nieuvenhuizen, Phys. Lett. {\bf B243}
 (1990) 245;\hfil\break
 E. Bergshoeff, A. Bilal and K.S. Stelle, CERN preprint TH 5924/90 and
 references therein.

 \item{[10]}
 P. Mathieu, Phys. Lett. {\bf B208} (1988) 101; J. Math. Phys. {\bf 29}
 (1988) 2499;\hfil\break
 I. Bakas, Phys. Lett. {\bf B219} (1989)\quad ; Com. Math. Phys.
 {\bf 123} (1989) 627;\hfil\break
 J.D. Smit, Com. Math. Phys. {\bf 128} (1990) 1.
 \vfill\eject

 \item{[11]}
 B. Kupershmidt, {\it Integrable and Superintegrable Systems} (World
 Scientific, 1990);\hfil\break
 A. Das, {\it Integrable Models} (World Scientific, 1989);\hfil\break
 M.A. Olshanetsky and A.M. Peremolov, Phys. Rep. {\bf 71} (1981)
 315;\hfil\break
 L.D. Fadeev and L.A. Takhtajan, {\it Hamiltonian Methods in the Theory
 of Solitons} (Springer, Berlin, 1987).

 \item{[12]}
 E.H. Saidi and M. Zakkari, Int. J. Mod. Phys. {\bf A6} (1991) 3151;
 {\bf A6} (1991) 3175.

 \item{[13]}
 W. Lerche, C. Vafa and N.P. Warner, Nucl. Phys. {\bf B324} (1989)
 427;\hfil\break
 C. Vafa and N.P. Warner, Phys. Lett. {\bf B218} (1989) 51.

 \item{[14]}
 E.H. Saidi and M.B. Sedra, Class. Quant. Gav.V10(1993)1937-1946; \hfil\break
 E.H. Saidi and M.B. Sedra, Int. Jour.Mod.Phys A6V9(1994)891; \hfil\break
 E.H. Saidi and M.B. Sedra, Mod. Phys. LettA V9 N34(1994)3163-3173;
\hfil\break
 M.B. Sedra, These de 3\`eme cycle (1993) LMPHE. Fac.Sciences Rabat

 \item{[15]}
 T. Inami and H. Kanno, Kyoto preprint YITP/K--929 (1991) and Nucl. Phys.
 {\bf B359} (1991) 201.

 \item{[16]}
 J. Evans and T. Hollowood, Nucl. Phys. {\bf B353} (1991) 529;\hfil\break
 H. Nohara, Annals of Physics {\bf 214} (1992) 1;\hfil\break
 S. Komata, K. Mohri and H. Nohara, Nucl. Phys. {\bf B359} (1991) 168.

 \item{[17]}
 V. Kac, Advances in Math. {\bf 26} (1977) 8;\hfil\break
 M. Scheunert, The Theory of Lie Superalgebras, Lecture Notes in Math.
 (1979).

 \item{[18]}
 A. Frappat, A. Sciarrio and P. Sorba, Com. Math. Phys. {\bf 88} (1983)
 63;\hfil\break
 D.A. Leites, M.V. Saveliev and V.V. Serganova, I HEP 85--81 (1985).

 \item{[19]}
 M.A. Olshanetsky, Com. Math. Phys. {\bf 88} (1983) 63;\hfil\break
 A.N. Leznov and M.V. Saveliev, Teor. Math. Fiz. {\bf 61} (1985)
 1056;\hfil\break
 V.A. Andreev, Teor. Math. Fiz. {\bf 72} (1988) 758.

 \item{[20]}
 See for instance\hfil\break
 M.F. de Groot, T.J. Hollowood and J.L. Miramontes, Com. Math. Phys.
 {\bf 145} (1992) 57 and references therein.

 \item{[21]}
 E.H. Saidi, in preparation.

 \item{[22]}
 C. Vafa and N. Warner, Phys. Lett. {\bf B218} (1989) 51;\hfil\break
 W. Lerche and N. Warner, preprint CALT 68 1703;\hfil\break
 N. Warner, Talk given at the Summer Workshop in High Energy Physics,
 ICTP, Trieste, (1992).
 \vfill\eject

 \item{[23]}
 S. Cecotti and C. Vafa, Nucl. Phys. {\bf B367} (1991) 359 and references
 therein;\hfil\break
 C. Vafa, Talk given at the Summer Workshop in High Energy Physics,
 ICTP, Trieste, (1992);\hfil\break
 T. Eguchi and S.K. Yang, Mod. Phys. Lett. {\bf A4} (1990) 1653.

 \item{[24]}
 P. Di Vecchia, J. Peterson and H. Zhang, Phys. Lett. {\bf B162} (1986)
 327.

 \item{[25]}
 M. Yu and H. Zheng, Nucl. Phys. {\bf B288} (1987) 275.

 \item{[26]}
 E. Ivanov and S. Krivonos, ``Superfield realizations of $N=2$ super $W_3$'',
 ICTP, Trieste, preprint No.IC/92/64 (1992).

 \item{[27]}
 T. Inami and H. Kanno, ``$N=2$ super $W$ algebras and generalized $N=2$
 super KdV hierarchies based on Lie superalgebras'', Kyoto preprint
 YITP/K--928 (1991).
 \vfill\eject
 \bye